\documentclass[lettersize,journal]{IEEEtran}
\usepackage{amsmath,amsfonts}
\usepackage{algorithmic}
\usepackage{algorithm}
\usepackage{array}
\usepackage[caption=false,font=footnotesize,labelfont=rm,textfont=rm]{subfig}
\usepackage{textcomp}
\usepackage{stfloats}
\usepackage{url}
\usepackage{verbatim}
\usepackage{graphicx}
\usepackage{cite}
\usepackage{stfloats}
\usepackage{placeins}
\usepackage{lipsum} 

\begin{document}

\title{A Homogeneous Graph Neural Network for Precoding and Power Allocation in Scalable Wireless Networks}

\author{Mingjun Sun,~\IEEEmembership{Graduate Student Member,~IEEE} Shaochuan Wu,~\IEEEmembership{Senior Member,~IEEE}, Haojie Wang, \\
Yuanwei Liu,~\IEEEmembership{Fellow,~IEEE}, Guoyu Li,~\IEEEmembership{Graduate Student Member,~IEEE}, and Tong Zhang.
\thanks{This work was supported by National Natural Science Foundation of China under Grant 62271167. \textit{(Corresponding author: Shaochuan Wu.)}}
\thanks{Mingjun Sun, Shaochuan Wu, Haojie Wang, Guoyu Li, and Tong Zhang are with the School of Electronics and Information Engineering, Harbin Institute of Technology, Harbin 150001, China (e-mail: sunmj@stu.hit.edu.cn; scwu@hit.edu.cn; \{22B905021, lgy\}@stu.hit.edu.cn; zhang7099@foxmail.com).}
\thanks{Yuanwei Liu is with the School of Electrical and Electronic Engineering, The University of Hong Kong, Hong Kong (e-mail: yuanwei@eee.hku.hk).}
}

\maketitle

\begin{abstract}
Deep learning is widely used in wireless communications but struggles with fixed neural network sizes, which limit their adaptability in environments where the number of users and antennas varies. To overcome this, this paper introduced a generalization strategy for precoding and power allocation in scalable wireless networks. Initially, we employ an innovative approach to abstract the wireless network into a homogeneous graph. This primarily focuses on bypassing the heterogeneous features between transmitter (TX) and user entities to construct a virtual homogeneous graph serving optimization objectives, thereby enabling all nodes in the virtual graph to share the same neural network. This ``TX entity'' is known as a base station (BS) in cellular networks and an access point (AP) in cell-free networks. Subsequently, we design a universal graph neural network, termed the information-carrying graph neural network (ICGNN), to capture and integrate information from this graph, maintaining permutation invariance. Lastly, using ICGNN as the core algorithm, we tailor the neural network's input and output for specific problem requirements and validate its performance in two scenarios: 1) in cellular networks, we develop a matrix-inverse-free multi-user multi-input multi-output (MU-MIMO) precoding scheme using the conjugate gradient (CG) method, adaptable to varying user and antenna numbers; 2) in a cell-free network, facing dynamic variations in the number of users served by APs, the number of APs serving each user, and the number of antennas per AP, we propose a universal power allocation scheme. Simulations demonstrate that the proposed approach not only significantly reduces computational complexity but also achieves, and potentially exceeds, the spectral efficiency (SE) of conventional algorithms.
\end{abstract}


\section{Introduction}
\IEEEPARstart{P}{recoding} and power allocation technologies are promising for enhancing wireless communication capacity by suppressing user interference.\cite{ref1,ref2}. However, these issues are often modeled as non-convex NP-hard problems\cite{ref3}, making them difficult to solve, as exemplified by the weighted sum rate (WSR) maximization problem\cite{ref4,ref5,ref6}. The weighted minimum mean square error (WMMSE) algorithm provides a closed-form expression for achieving local optimal solutions and is widely used as a benchmark\cite{ref5}. Nevertheless, the iterative solving process and high-dimensional matrix inversion operations of the algorithm result in significant computational delays. These delays become particularly problematic in dynamically changing wireless communication environments. Such as when the number of users or antennas changes, requiring the algorithm to reiterate and thereby making it unable to meet real-time requirements. Therefore, designing an efficient and versatile algorithm has become urgently necessary.

\subsection{Prior Works}
Deep learning technology offers new solutions for balancing algorithm performance and complexity. In multi-user multiple-input single-output (MU-MISO) precoding scenarios, Sun et al. in \cite{ref7} treated iterative optimization algorithms as an unknown nonlinear function and demonstrated that it could be approximated by a deep neural network (DNN). They trained the DNN using supervised learning with labeled datasets derived from the WMMSE algorithm. However, the performance upper bound of this scheme is limited by WMMSE. In \cite{ref8}, the authors proposed a method to construct the loss function by linking it to SE and using unsupervised learning techniques to overcome the performance limitations of WMMSE.
To simplify the output layer dimension of the network, the authors of \cite{ref9} used the expert knowledge such as uplink-downlink duality and known optimal solution structures to design the network to output only two power allocation factors, facilitating faster convergence. 
Additionally, some studies focused on enhancing the interpretability of neural networks, proposing the concept of deep unfolding neural networks \cite{ref11,ref13,ref14}. The core idea is to map each iteration of the iterative algorithms to each layer of the neural network and introduce trainable modules (at parameter level or network level) to replace the high-latency computation process of traditional algorithms. The authors in \cite{ref11} investigated the power allocation problem for single-antenna transceivers in ad hoc networks. They intorduced a network-level trainable module, i.e., graph neural networks (GNNs), to establish connections between the underlying wireless network and the auxiliary variables of WMMSE, while preserving the key structure of the WMMSE algorithm, thereby accelerating algorithm convergence. 
In \cite{ref13}, the iterative algorithm induced deep unfolding neural network (IAIDNN) algorithm was proposed for the MU-MIMO beamforming scenario, which used a parameter-level trainable module to compensate for truncation errors caused by the first-order Taylor expansion approximation of matrix inverse operations, effectively reducing computational complexity. In \cite{ref14}, the authors employed projected gradient descent (PGD) instead of the Lagrange multiplier algorithm for acquiring precoding vectors in MU-MISO scenarios, avoiding complex operations such as matrix inversion, eigenvalue decomposition, and bisection methods. The step size of gradient descent was set as a trainable parameter, reducing computational complexity while achieving performance similar to that of WMMSE.

Previous studies have effectively reduced computational complexity and somewhat met the low-latency requirements of communication systems. However, the universality of neural networks remains to be further explored.
When the scale of the wireless network changes, it necessitates redesigning and retraining neural networks, which fails to meet real-time requirements.
Some studies addressed the above issues by modeling wireless networks as graphs and employing either homogeneous or heterogeneous GNNs, based on the graph's homogeneity or heterogeneity, to accomplish wireless resource allocation tasks effectively \cite{ref38, ref39, ref19, ref37, ref40, ref41}.
In \cite{ref38}, the authors modeled the power allocation problem in cellular networks as a heterogeneous graph, with BS antennas and users represented as two types of nodes. They employed a heterogeneous GNN to address this problem, effectively handling variations in the number of users. Building on this, \cite{ref39} further enhanced the structure of heterogeneous GNNs by enabling feature updating on the edges of heterogeneous graphs, thereby broadening their applicability to more general scenarios.
The studies in \cite{ref19} and \cite{ref37} investigated beamforming and link scheduling in device to device (D2D) networks, respectively. Since transceivers exist in pairs, each transceiver pair is treated as a single node, allowing for the construction of a homogeneous graph to effectively handle variations in the number of D2D pairs.
For the multi-user downlink beamforming problem, the author in \cite{ref40} followed an approach similar to that of \cite{ref38}, treating each antenna of a centralized BS as a distributed node. Distinct message-passing networks were designed for BS antenna nodes and user nodes, enabling adaptability to changes in the numbers of BS antennas and users. Meanwhile, reference \cite{ref41} modeled the MU-MISO scenario as a homogeneous graph, where the communication links between the BS and users were treated as nodes without assigning features to the edges. The authors incorporated an attention mechanism into the GNN to enhance its learning capability, directly mapping the channel and precoding vectors. This approach enabled the model to adapt effectively to changes in the number of users.

The above studies enable network scalability\footnote{In this work, the terms ``scalability'' or ``scalable'' specifically refer to the capability of neural network-based precoding algorithms or power allocation methods to adapt to variations in wireless network scale, such as changes in the number of users. This does not imply algorithmic adaptability under asymptotic regimes where the scale approaches infinity, as such scenarios would result in unbounded growth in both the number of graph nodes and the feature dimensions of nodes/edges, which falls outside the scope of our consideration.} primarily because the number of trainable parameters in the designed neural networks is independent of dynamically changing variables, such as the number of BS antennas and users. Specifically, in the GNN structure, the input and output layer dimensions of the feature aggregation or message-passing networks are fixed, unaffected by the number of nodes in the graph. This design allows the model to adapt to graphs of varying sizes effectively.

Recently, large artificial intelligence models, represented by ChatGPT, have sparked an incredible revolution. By undergoing extensive training in specific domains, they can adapt to a wide range of downstream tasks with only few-shot or even zero-shot learning. This suggests that universal intelligence in the mobile communications field might need to break away from the limitations of task-specificity\cite{ref15}. Accordingly, we aim to explore a preliminary, unified architecture that, from constructing wireless network graphs to designing GNNs, could be adaptable to resource allocation problems across various scenarios.

\subsection{Motivations and Contributions} 
Compared to complex social networks, wireless communication network models have no multi-hop neighboring nodes. Homogeneous GNNs possess the ability to learn their features, which has also been demonstrated \cite{ref19,ref26}. This paper is related to \cite{ref19}, which presents a modeling approach for D2D network graphs and has been widely adopted due to the simplicity of the proposed wireless channel graph convolution network (WCGCN) and its rigorous theoretical proof. Building on this foundation, we further extend homogeneous graph modeling to broader scenarios and improve the neural network design. First, we propose a general method for constructing wireless networks as homogeneous virtual graphs. Next, we develop an information-carrying GNN (ICGNN) that enables feature aggregation and updating. To validate this architecture, we then examine the MU-MIMO downlink precoding problem in cellular networks and the power allocation problem in cell-free networks. In previous studies \cite{ref19,ref39,ref40,ref41}, the beamforming problem is typically formulated under the assumption that user devices are equipped with a single antenna. This simplifies the dimensionality of the optimization variables and facilitates GNN training. However, research on scenarios involving multi-antenna users remains limited. In addition, this paper considers the generalization problem caused by changes in the number of receiving antennas due to variations in mobile terminal types. Furthermore, we also validate the unified nature of the proposed framework in the power allocation problem within cell-free networks. Specifically, we design a power allocation scheme for a cell-free network that can accommodate variations in three parameters: the number of AP antennas, the number of APs serving users, and the number of users served by each AP.


\begin{figure*}[!t]
\vspace{-10pt}
\centering
\includegraphics[scale=0.56]{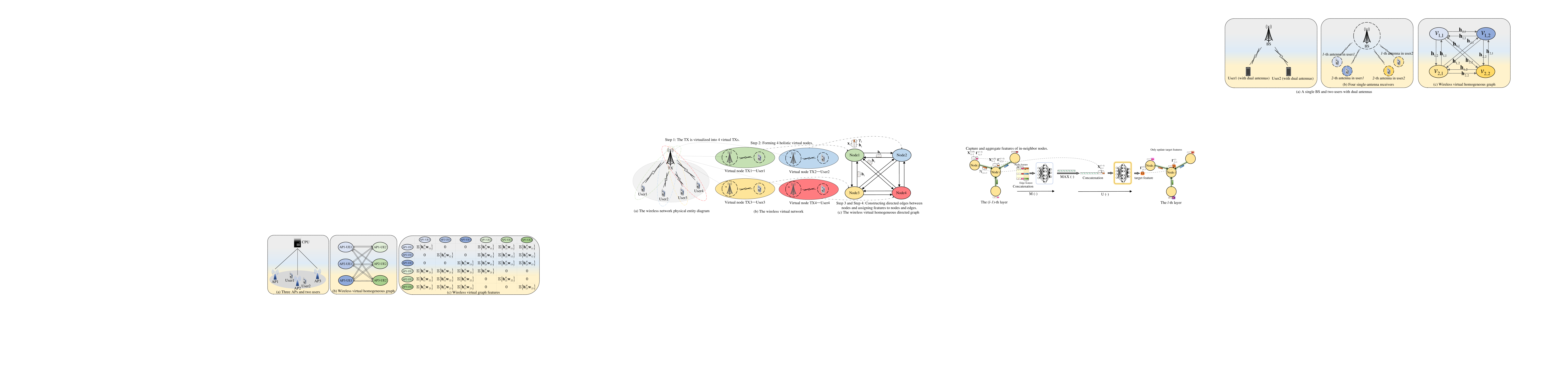}
\caption{Abstracting wireless networks as wireless virtual homogeneous graph.}
\vspace{-5pt}
\label{fig:1}
\vspace{-15pt}
\end{figure*}

The specific contributions of this paper are summarized as follows.
\begin{itemize}
\item{We propose a method for constructing wireless networks as universal homogeneous underlying graphs. Rather than emphasizing the heterogeneous characteristics between TX and user entities, this approach abstracts a homogeneous virtual graph suited for optimization objectives. It can serve as a reference for modeling other wireless networks composed of fully-digital TXs and users as homogeneous graphs. Moreover, it has been demonstrated that optimization problems established on the graph, such as precoding and power allocation, possess permutation invariance properties.}
\item{A node-wise graph neural network, namely ICGNN, has been designed, which is independent of the number of nodes in the virtual graph. It can adapt to graphs of varying sizes while maintaining permutation invariance. During the updating of node features, we only update the target features while keeping the channel features constant, which helps to reduce the network size. The target features and channel features correspond to the attributes associated with the optimization variables and channel state information (CSI), respectively.}
\item{An ICGNN-based MU-MIMO precoding method, adaptable to cellular networks with varying numbers of users and user antennas, has been developed. Initially, this method converts the MU-MIMO optimization problem into a MU-MISO problem, subsequently deriving the precoding vectors using the optimal solution structure. Additionally, we introduce a matrix-inverse-free precoding recovery method based on the CG method, which further reduces computational complexity. Simulation results demonstrate that this approach achieves performance comparable to, or even surpassing, that of the WMMSE algorithm.}
\item{We designed a universal power allocation scheme for cell-free networks to further verify the effectiveness of the proposed architecture. This scheme can accommodate scenarios with variable numbers of APs, users, and AP antennas. Simulation results show that this scheme delivers performance nearly identical to WMMSE while also significantly reducing computational complexity.}
\end{itemize}

The remainder of this paper is organized as follows. Section II A introduces how to construct a universal wireless virtual graph and demonstrates the permutation invariance of graph optimization problems.  Section II B presents the specific structure of the proposed ICGNN and its advantages. Section III discusses the MU-MIMO precoding scheme in cellular networks, and the power allocation scheme in cell-free networks is given in Section IV. Simulation results are provided in Section V and conclusions are drawn in Section VI.

\textit{Notations:} We denote matrices, vectors and scalar by bold-face upper-case,  bold-face lower-case and italic letter. $\left(\cdot\right)^T$, $\left(\cdot\right)^H$, $\left(\cdot\right)^{-1}$ and $\operatorname{Tr}\left(\cdot\right)$ denote the transpose, the Hermitian transpose, the inverse and the trace operation, respectively.
We use $\left(\cdot\right)^{(i)}$ to denote the variable after the $i$-th iteration update, $\mathbb{C}^{M \times N}$ to represent an $M \times N$ complex matrix, $|\cdot|$ to indicate the cardinality of a set, and $\|\cdot\|_{\rm F}$ denote the Frobenius norm of a matrix. 

\section{wireless virtual graph and graph neural network}

\begin{figure*}[!t]
\vspace{-10pt} 
\centering
\includegraphics[scale=0.72]{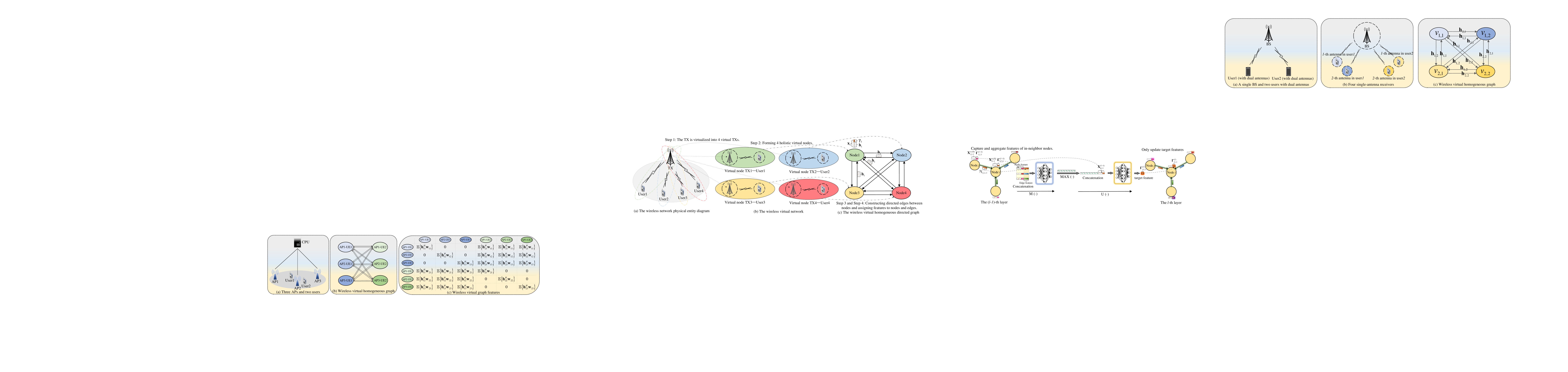}
\caption{The message passing and feature updating processes of the ICGNN network.}
\label{fig:2}
\vspace{-10pt} 
\end{figure*}

\subsection{Constructing a Wireless Virtual Graph}
We first give the following definition about graphs.

\textit{Definition 1 (About Graph): 
A graph consists of nodes and edges. It is homogeneous if it contains only one type of node and edge, and heterogeneous if it has multiple types. If edges between nodes are bidirectional, the graph is undirected; if they have a defined direction, it is directed.}

Generally, in wireless networks, as shown in Fig. 1(a), the TX and users can be considered as nodes, and the direct communication channels between them as edges, thus forming an undirected graph\footnote{For example, if we consider the downlink related problem, then the edge between the TX and the user node only indicates the existence of a direct communication link between the two without considering the true direction of data transmission, and is therefore defined as an undirected edge.}. However, the TX and user entities exhibit distinct heterogeneous characteristics; for example, the user-side features may include receiver noise power and priority coefficients, while TX-side features are mainly concerned with resource allocation. Consequently, some literature suggests modeling this as a heterogeneous undirected graph and designing specialized feature aggregation and merging neural networks for different types of nodes\cite{ref38, ref40}. However, we propose that constructing a wireless network graph can leverage prior knowledge to overlook the heterogeneous characteristics between physical entities and build a virtual homogeneous directed graph that serves optimization problems, as illustrated in Fig. 1(c). This allows all nodes in the homogeneous graph to share the same neural network, simplifying the network size. The detailed construction process of the homogeneous directed graph is given next. 
\begin{itemize}
\item{\textit{Step 1}, the TX is virtualized into multiple virtual TX nodes, with the number matching the number of users they serve. Assuming that the users $1$ through $4$ are currently accessing the network, the TX is virtualized into $4$ virtual TXs.}
\item{\textit{Step 2}, these virtual TXs and their served users are considered as a single holistic virtual node (VN), as shown in Fig. 1(b). }
\item{\textit{Step 3}, based on the interference relationships between users, the VNs are interconnected to construct virtual directed edges. The reason for having directed edges is that the interference links between any two users differ in how they interfere with each other. For example, the interference from User $1$ to User $2$ comes from the data sent by the TX to User $1$, which is received by User $2$ through the channel from the TX to User $2$. Similarly, the interference link from User $2$ to User $1$ is through the channel from the TX to User $1$. Therefore, the edges between any two nodes are set as directed edges. }
\item{\textit{Step 4}, using prior knowledge, features are assigned to virtual nodes and edges.}
\end{itemize}
In wireless communication networks, both the precoding matrix and the power allocation factor are directly determined by the CSI. Therefore, we assign the features related to the CSI (such as CSI itself or others) to all edges, referred to as channel features. As for the node features, they are concatenated with other target features after the channel features, based on the specific problem requirements. Taking the Node $1$ as an example, the feature of the edges pointing to Node $1$ from other nodes can be defined as $\mathbf{h}_1$, which denotes the CSI from the TX to User $1$. This definition is due to the fact that Node $1$ receives all interfering signals through $\mathbf{h}_1$.
The feature $\mathbf{x}_1\in \mathbb{C}^{d_1}$ of Node $1$ is formed by concatenating the channel feature $\mathbf{h}_1\in \mathbb{C}^{d_2}$ with the target feature $\boldsymbol{\gamma}_1\in \mathbb{C}^{d_3}$, i.e., $\mathbf{x}_1=(\mathbf{h}_1, \boldsymbol{\gamma}_1)$, as shown in Fig. 1(c). The operator $(\cdot, \cdot)$ denotes concatenation. Let $d_1$, $d_2$, and $d_3$ represent the dimensions of the corresponding features, where $d_1 = d_2 + d_3$. In wireless resource allocation problems, the target features are typically set as variables to be optimized, such as precoding matrices or power allocation factors. 

Therefore, the proposed wireless virtual graph (WVG) can be represented as $\mathit{G}=(\mathit{V},\mathit{E})$, where $\mathit{V}=\left\{\mathit{v}_{1}, \mathit{v}_{2}, \ldots, \mathit{v}_{|V|} \right\}$ denotes the set of VNs, and $|\cdot|$ indicates the cardinality of a set, while $\mathit{E}=\left\{\mathit{e_{ij}}, 1\leq \forall i,j \leq |V| \right\}$ represents the set of directed edges, where $\mathit{e_{ij}}$ is a directed edge from $\mathit{v}_{i}$ to $\mathit{v}_{j}$.
The node feature matrix is defined as $\mathbf{X} \in \mathbb{C}^{|V|\times \mathit{d_1}}$, where $\mathbf{X}_{(i,:)}=\mathbf{x}_i=(\mathbf{h}_i, \boldsymbol{\gamma}_i)$ represents the feature vector of $\mathit{v}_{i}$. Here, $\mathbf{h}_i\in \mathbb{C}^{d_2}$ and $\boldsymbol{\gamma}_i\in \mathbb{C}^{d_3}$ represent the channel feature and target feature of $\mathit{v}_{i}$, respectively. The features of the edges can be represented as a tensor $\mathcal{H}\in \mathbb{C}^{|V|\times |V| \times \mathit{d_2}}$, where $\mathcal{H}_{(i,j,:)}$ represents the feature vector of edge $\mathit{e_{ij}}$. If $i = j$, $\mathcal{H}_{(i,j,:)}= \mathbf{0}\in \mathbb{C}^{d_2}$, since there are no self-connecting edges at a VN. To make it more intuitive, we extract the target feature matrix from the node feature matrix $\mathbf{X}$ and redefine it as $\mathbf{\Gamma}\in \mathbb{C}^{|V|\times \mathit{d_3}}$. Here, $\mathbf{\Gamma}_{(i,:)}=\boldsymbol{\gamma}_i$ represents the optimization variable corresponding to $\mathit{v}_{i}$ in a graph optimization problem. Additionally, we extract the channel feature matrix from $\mathbf{X}$ and define it as $\mathbf{X}^{rc}\in \mathbb{C}^{|V|\times \mathit{d_2}}$. Moreover, the row vector $\mathbf{X}^{rc}_{(i,:)}=\mathbf{x}^{rc}_i$ corresponds to $\mathbf{h}_i$ for the VN $\mathit{v}_{i}$.

At this point, we can formulate the following graph optimization problem on the WVG,
\begin{equation}
\label{eqa:1}
\begin{aligned}
\mathcal{P}_1: \max_{\Gamma} & \quad \mathit{g}(\boldsymbol{\alpha},\mathbf{\Gamma}, \mathbf{X}^{rc}, \mathcal{H}) \\
\text {s.t.} & \quad \mathit{C}(\mathbf{\Gamma}, \mathbf{X}^{rc}, \mathcal{H})\leq 0,
\end{aligned}
\end{equation}
where $g(\cdot,\cdot,\cdot)$ and $\mathit{C}(\cdot, \cdot, \cdot)$ are the objective function and constraint function, respectively. They are composed of symmetric operations such as summation, maximum, minimum, inner product, etc. Additionally, $\boldsymbol{\alpha}=[\alpha_1,\alpha_2,\ldots,\alpha_{|V|}]^T\in \mathbb{R}^{|V|}$ denotes the weight factor corresponding to each VN.

The function $g(\cdot,\cdot,\cdot)$ and $\mathit{C}(\cdot, \cdot, \cdot)$ can be proven to follow permutation invariance. To illustrate this point, we first define a permutation operation $\pi$ on VNs, which represents a bijective function from the set $V$ to itself, i.e., $\pi: V\to V$. 
In the wireless virtual graph $\mathit{G}$, a permutation of nodes, denoted by ``$\pi \star \cdot$'', reorders the elements as follows: for $\boldsymbol{\alpha}$, the permutation is defined by $(\pi \star \boldsymbol{\alpha})_{\pi(i)}= \alpha_{i}$, meaning that the elements of $\boldsymbol{\alpha}$ are permuted according to $\pi$. For $\mathbf{X}^{rc}$, it is defined by $(\pi \star \mathbf{X}^{rc})_{(\pi(i),:)}= \mathbf{X}^{rc}_{(i,:)}$, which means that the rows of $\mathbf{X}^{rc}$ are permuted according to $\pi$. For $\mathbf{\Gamma}$, it is similarly defined by $(\pi \star\mathbf{\Gamma})_{(\pi(i),:)}=\mathbf{\Gamma}_{(i,:)}$, also indicating a row permutation. For $\mathcal{H}$, the permutation is defined by $(\pi \star \mathcal{H})_{(\pi(i),\pi(j),:)}=\mathcal{H}_{(i,j,:)}$, which rearranges both the rows and columns. In all these cases, the graph's features corresponding to the permuted indices remain identical to the features associated with the original indices. This ensures that the structure and features of the graph remain unchanged under permutation. We say that the WVG $G$ is isomorphic before and after the permutation.

\textit{Proposition 1 (Permutation Invariance): The objective function $g(\cdot,\cdot,\cdot,\cdot)$ and its constraint $C(\cdot,\cdot,\cdot)$ satisfy the property $\mathit{g}(\pi \star \boldsymbol{\alpha},\pi \star \mathbf{\Gamma}, \pi \star\mathbf{X}^{rc}, \pi \star\mathcal{H})= \mathit{g}(\boldsymbol{\alpha},\mathbf{\Gamma}, \mathbf{X}^{rc}, \mathcal{H})$ and $\mathit{C}(\pi \star\mathbf{\Gamma}, \pi \star\mathbf{X}^{rc}, \pi \star\mathcal{H})=\mathit{C}(\mathbf{\Gamma}, \mathbf{X}^{rc}, \mathcal{H})$, for any permutation operator $\pi$.}

\textit{Reamrk 1:} 
A detailed proof is provided in Appendix A.

\subsection{Information Carrying Graph Neural Network}

First, we introduce the basic framework of message passing neural networks (MPNNs) \cite{ref18} that operate on a directed graph $G'$. For simplicity, we define the features of node $i$ and its neighboring node $j$ as $x'_i \in \mathbb{C}$ and $x'_j \in \mathbb{C}$, respectively. The feature of the edge directed from node $j$ to node $i$ is defined as $e'_{ji} \in \mathbb{C}$.
In MPNNs, the forward inference process can be divided into two phases: the message passing phase and the node feature update phase. The message passing phase is responsible for capturing and aggregating the features of in-neighbor nodes, represented by the message function $\mathrm{M}_t(\cdot)$. The node feature update phase updates the features of the current node based on the information gathered during the message passing phase, represented by the node update function $\mathrm{U}_t(\cdot)$. Here, $t$ refers to the $t$-th iteration of forward inference, which corresponds to the $t$-th layer of the MPNN. These two phases can be represented as shown in (2) and (3).
\begin{equation}
\vspace{-5pt}
\label{eqa:2}
m_i^{t+1}= \underset{j \in \mathcal{N}(i)}{\operatorname{AGGREGATE}}\{\mathrm{M}_t((x'^{(t)}_j,e'^{(t)}_{ji}))\},
\end{equation}

\begin{equation}
\vspace{-1pt}
\label{eqa:3}
x'^{(t+1)}_i=\mathrm{U}_t((x'^{(t)}_i,m_i^{t+1})),
\end{equation}
where $\mathcal{N}(i)$ and $m_i^{t+1}\in \mathbb{C}$ denote the neighbors\footnote{Unless otherwise specified, all neighbors referred to in the following text are in-neighbors.} of node $i$ and the output of the message passing phase at the $t$-th layer of the MPNN, respectively. Additionally, the AGGREGATE represents an aggregation function for the features of neighboring nodes. The aggregation function must remain invariant to the ordering of the input data, and it should be able to handle a variable number of input parameters. Common aggregation operations include elementwise summation, mean, and maximum\cite{ref21}. 

\begin{figure*}[b]
\vspace{-10pt} 
\centering
\hrule 
\vspace{-1pt} 
\begin{subequations}
\renewcommand{\theequation}{\theparentequation\alph{equation}} 
\begin{align}
\mathbf{\Gamma}_{(i,:)}^{(1)}&=\mathrm{U}\left(\left(\mathbf{X}_{(i,:)}^{(0)}, \underset{j \in \mathcal{N}(i)}{\operatorname{AGGREGATE}}\left\{\mathrm{M}\left(\left(\mathbf{X}_{(j,:)}^{(0)}, \mathcal{H}_{(j, i,:)}\right)\right)\right\}\right)\right), \\
& \cdots, \nonumber\\ 
\mathbf{\Gamma}_{(i,:)}^{(l)}& =\mathrm{U}\left(\left(\left(\mathbf{X}_{(i, :)}^{rc(0)}, \mathbf{\Gamma}_{(i,:)}^{(l-1)}\right), \underset{j \in \mathcal{N}(i)}{\operatorname{AGGREGATE}}\left\{\mathrm{M}\left(\left(\left(\mathbf{X}_{(j, :)}^{rc(0)}, \mathbf{\Gamma}_{(j,:)}^{(l-1)}\right), \mathcal{H}_{(j, i, :)}\right)\right)\right\}\right)\right).
\end{align}
\end{subequations}
\end{figure*}

Below, we provide a detailed description of the specific architecture of the ICGNN. The output of the $l$-th layer for the VN $\mathit{v}_{i}$ can be represented by (4a) and (4b) as shown at the bottom of this page, and this process is clearly illustrated in Fig. 2. Here, $\mathrm{M}(\cdot)$ and $\mathrm{U}(\cdot)$  refer to the message passing network and the node feature update network, respectively, both of which are composed of fully connected layers. The detailed structures of these networks will be discussed in Sections III and IV.
The initial feature $\mathbf{X}_{(i,:)}^{(0)}\in \mathbb{C}^{d_1}$ of the VN $\mathit{v}_{i}$ is given by $\mathbf{X}_{(i,:)}^{(0)}=\left(\mathbf{X}_{(i, :)}^{rc(0)}, \mathbf{\Gamma}_{(i,:)}^{(0)}\right)$, where $\mathbf{X}_{(i, :)}^{rc(0)}\in \mathbb{C}^{d_2}$ and $\mathbf{\Gamma}_{(i,:)}^{(0)} \in \mathbb{C}^{d_3}$ represent the initial channel and target feature of $\mathit{v}_{i}$. Additionally, $\mathbf{\Gamma}_{(i,:)}^{(l)}$ denotes the $l$-th update of the target feature of $\mathit{v}_{i}$.

The forward inference process of the ICGNN differs from that of the MPNN.
Its uniqueness lies in the fact that during the update of node features, $\{\mathbf{X}_{(i, :)}^{rc(0)},\forall i\}$ remains unchanged, with only the target features in $\{\mathbf{\Gamma}_{(i,:)}^{(0)},\forall i\}$ being updated. 
This design is based on the following reasons: 
(i) In the WVG, the VNs are directly connected, so only one message passing phase is needed to gather information about all neighboring VNs. Unlike traditional MPNNs, which require $t$ iterations to obtain information from $t$-hop neighbors. In other words, a single execution of $\mathrm{M}(\cdot)$ (one layer of ICGNN) is sufficient to acquire all the information needed to update the target features of the VNs. There's no need to update the channel features of the VNs to include information from neighboring VNs and then wait for the ICGNN to propagate through $t$ layers to integrate $t$-hop neighbor information for target feature updates. This process would be redundant. 
(ii) We still design the ICGNN network as a multi-layer architecture, typically selecting two to three layers based on actual requirements. The purpose is not to obtain multi-hop neighbor information but rather because the target features updated by the ICGNN are closer to the optimal solution compared to the initial manually assigned strategy. However, these valuable values are unattainable manually. Combining them with wireless channel features and re-entering them into the ICGNN network facilitates faster convergence.
(iii) By updating only the target features of the VNs, the output layer size of the node feature update network $\mathrm{U}(\cdot)$ can be reduced from $d_1$ to $d_3$, where $d_1$ is directly related to $N_t$, with $N_t$ being the number of BS antennas.

In ICGNN, the aggregation function is set to the $\mathrm{MAX}(\cdot)$ function because this function embeds a non-linear relationship into the GNN structure, which is advantageous for GNN extrapolation to unseen graph sizes\cite{ref20,ref24}.
In \cite{ref20}, the authors use the max-degree problem in graph theory as a case study, aligning the MLP module within a GNN to approximate linear sub-functions of the target function, while employing the $\mathrm{MAX}(\cdot)$ aggregation to capture the necessary nonlinearity. Both theoretical and empirical results suggest that such a design can enhance the extrapolation capability of GNNs. Similarly, the authors in \cite{ref24} demonstrate that the $\mathrm{MAX}(\cdot)$ function improves GNN extrapolation performance across various graph tasks. While the conclusions in \cite{ref20} and \cite{ref24} are based on specific assumptions, they provide valuable intuition for the design of GNN architectures in more general settings. Additionally, the authors in \cite{ref25} demonstrate that GNNs utilizing the $\mathrm{MAX}(\cdot)$ aggregation function maintain high output stability when the graph input features are subjected to slight disturbances, indicating that GNNs exhibit good robustness under imperfect CSI conditions. This has also been validated in the application of ad hoc networks by \cite{ref26}, where GNNs continue to deliver effective power allocation performance even in the absence of certain links.

In our case, the WSR maximization problem can be viewed as an unknown nonlinear mapping from the CSI to the optimal solution of the optimization variables. Inspired by the insights from \cite{ref20}, we adopt the ICGNN to fit this nonlinear target function may possess certain extrapolation capabilities.
This ability is particularly well-suited for wireless network environments, as it effectively handles the dynamic changes in the number of users accessing. 



After presenting the detailed design of the ICGNN, we demonstrate that ICGNN possesses the same permutation invariance properties as graph optimization problems.

\textit{Proposition 2 (Permutation Invariance in ICGNN): Considering the input-output mapping relationship of ICGNN as $\psi:\left(\mathbf{X}^{(0)}, \mathcal{H}\right) \mapsto \mathbf{X}^{(L)}$, we define the objective function on the graph as $g\left(\boldsymbol{\alpha},\psi\left(\mathbf{X}^{(0)}, \mathcal{H}\right),\mathcal{H}\right)=g\left(\boldsymbol{\alpha},\mathbf{X}^{(L)}, \mathcal{H}\right)$. Here, $\mathbf{X}^{(L)}=\left(\mathbf{X}^{rc(0)}, \mathbf{\Gamma}^{(L)}\right)$, where $\mathbf{X}^{(0)}$ and $\mathcal{H}$ represent the input node feature matrix and the edge feature tensor, respectively. $\mathbf{X}^{(L)}$ denotes the node feature matrix output by the ICGNN at the $L$-th layer. We have $g\left(\pi \star\boldsymbol{\alpha},\psi\left(\pi \star \mathbf{X}^{(0)}, \pi \star \mathcal{H}\right),\pi \star\mathcal{H}\right)=g\left(\pi \star\boldsymbol{\alpha},\pi \star \mathbf{X}^{(L)},\pi \star\mathcal{H}\right)=g\left(\boldsymbol{\alpha},\mathbf{X}^{(L)},\mathcal{H}\right)=g\left(\boldsymbol{\alpha},\psi\left(\mathbf{X}^{(0)}, \mathcal{H}\right),\mathcal{H}\right)$ for any permutation operator $\pi$.}

\textit{Remark 2:} Please refer to Appendix B for a detailed proof.

\begin{figure*}[!t]
\vspace{-10pt} 
\centering
\includegraphics[scale=0.57]{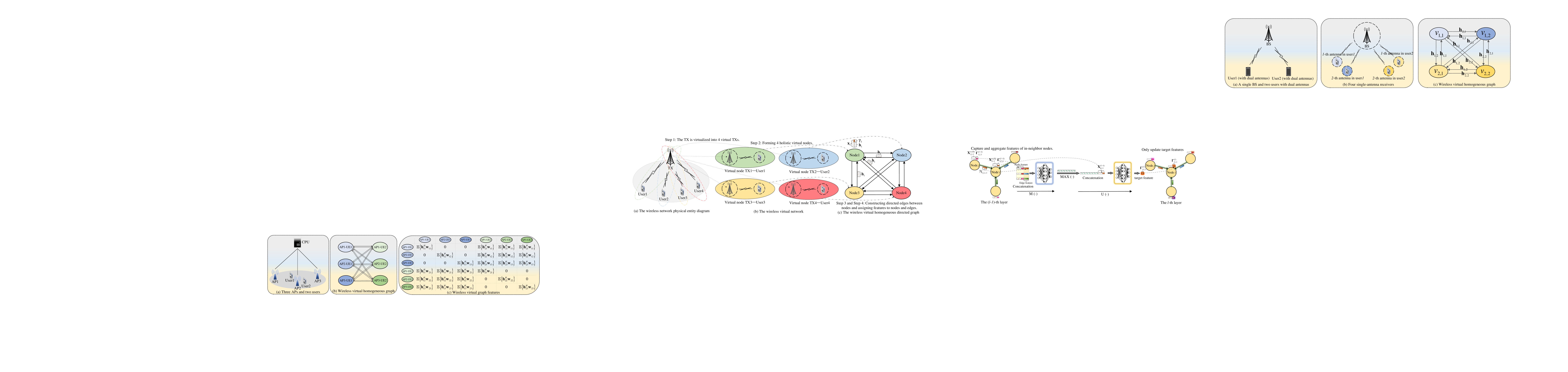}
\vspace{-10pt}
\caption{Wireless virtual graph of a MU-MIMO system.}
\label{fig:3}
\vspace{-15pt} 
\end{figure*}
\vspace{-1pt} 

\section{A Scalable Matrix-Inverse-Free MU-MIMO Precoding Algorithm in cellular scenario}
\subsection{System Model and Problem Formulation}
We consider a single-cell downlink MU-MIMO system, where the BS has $\mathit{N_{t}}$ transmit antennas serving $\mathit{K}$ users, each with $\mathit{N_{r}}$ receive antennas. The channel between user $\mathit{k}$ and the BS is $\mathbf{H}_{k} \in \mathbb{C}^{N_{t} \times N_{r}}$, and the BS uses a precoding matrix $\mathbf{V}_{k} \in \mathbb{C}^{N_{t} \times d_{k}}$ to process signals for user $\mathit{k}$, where $d_k$ is the number of data streams. The received signal for user $\mathit{k}$ is given by
\begin{equation}
\label{eqa:5}
\mathbf{y}_k=\underbrace{\mathbf{H}_k^H \mathbf{V}_k \mathbf{s}_k}_{\text{desired signal of user }k}+\underbrace{\sum_{m=1, m\neq k}^K\mathbf{H}_k^H \mathbf{V}_m \mathbf{s}_m}_{\text{inter-user interference}}+\mathbf{n}_k,\forall k,
\end{equation}
where $\mathbf{s}_k\in \mathbb{C}^{d_{k} \times 1} \sim \mathcal{CN}(\mathbf{0},\mathbf{I})$ represents the transmitted data for user $\mathit{k}$, and $\mathbf{n}_k \sim \mathcal{CN}(\mathbf{0},\sigma_k^2\mathbf{I})$ denotes the receiver's additive Gaussian white noise. Without loss of generality, we assume $d_k=N_r, \forall k$ in the following discussion. 

Then, the SE of user $k$ can be defined as $R_k$ in (6). 
\begin{equation}
\label{eqa:6}
R_k = \log_2 \operatorname{det}\left(\mathbf{I} + \mathrm{SINR}_k\right),
\end{equation}
where $\mathrm{SINR}_k$ denotes the signal to interference noise ratio (SINR) at user $\mathit{k}$, as formulated in (7).
\begin{equation}
\label{eqa:7}
\mathrm{SINR}_k = \mathbf{H}_k^H \mathbf{V}_k \mathbf{V}_k^H \mathbf{H}_k \left(\sum_{m \neq k} \mathbf{H}_k^H \mathbf{V}_m \mathbf{V}_m^H \mathbf{H}_k + \sigma_k^2 \mathbf{I}\right)^{-1}.
\end{equation}

The primary objective of downlink precoding in a MU-MIMO system is to design the precoding matrix $\mathbf{V}_k$ to maximize the SE. This optimization problem can be formulated as a WSR maximization problem subject to the total transmit power constraint:
\vspace{-5pt} 
\begin{equation}
\label{eqa:8}
\vspace{-5pt} 
\begin{aligned}
\mathcal{P}_2: \max_{\left\{\mathbf{V}_k\right\}} & \sum_{k=1}^K \alpha_k R_k \\
\text {s.t.} & \sum_{k=1}^K \operatorname{Tr}\left(\mathbf{V}_k \mathbf{V}_k^H\right) \leq P_t,
\end{aligned}
\end{equation}
where $\mathit{P}_t$ represents the total power budget of the BS, and $\alpha_k$ reflects the priority of user $\mathit{k}$. 
 The problem $\mathcal{P}_2$ is a typical non-convex NP-hard problem, for which a local optimum solution can be obtained using the WMMSE algorithm\cite{ref5}.

\subsection{Wireless Virtual Graph and Graph Optimization Problem}
In real scenarios, the number of users and their antennas ($K$ and $N_r$) served by the BS can vary over time. Thus, we aim to design a method that adapts to these changes.
Since each user has $N_r$ receive antennas, constructing a WVG with the user as the basic unit directly ties the dimensionality $d_1$ of each VN's features to $N_r$, such as flattening the channel matrix $\mathbf{H}_{k}$ and the precoding matrix $\mathbf{V}_{k}$ as the channel feature and the target feature respectively, i.e., $d_1=d_2 +d_3= N_tN_r+N_tN_r=2N_tN_r$. This makes the input and output layer sizes of the node-wise neural network dependent on $N_r$. To avoid redesigning and retraining the network when $N_r$ changes, we construct the WVG with the receive antenna as the smallest unit. For instance, each user with multiple antennas is regarded as multiple single-antenna receivers, illustrated in Fig. 3(a) and Fig. 3(b). Taking a single BS and two users with dual antennas as example, we construct the WVG as shown in Fig. 3(c), using the method in Section II A. This approach ensures that changes in the number of users or antennas only affect the total number of VNs in the WVG, without altering feature dimensionality, enabling the ICGNN to generalize effectively.

We further detail the specific form of the signal received by user $k$ on any receive antenna $i$. First, Equation (5) can be expressed as follows,
\vspace{-6pt} 
\begin{equation}
\vspace{-6pt} 
\label{eqa:9}
\begin{aligned}
\mathbf{y}_k&=\mathbf{H}_k^H\sum_{m=1}^K \mathbf{V}_m \mathbf{s}_m+\mathbf{n}_k,\forall k\\
&=\left[\mathbf{h}_{1,k}, \mathbf{h}_{2,k},\ldots, \mathbf{h}_{N_r,k}\right]^H \cdot \\
&\sum_{m=1}^K \left[\mathbf{v}_{1,m}, \mathbf{v}_{2,m},\ldots, \mathbf{v}_{N_r,m}\right] \mathbf{s}_m+\mathbf{n}_k,\forall k,
\end{aligned}
\end{equation}
where $\mathbf{H}_{k}=\left[\mathbf{h}_{1,k}, \mathbf{h}_{2,k},\ldots, \mathbf{h}_{N_r,k}\right]$ with each $\mathbf{h}_{i,k}\in \mathbb{C}^{N_{t} \times 1}$ represents the channel from the $i$-th antenna of user $k$ to the BS, and $\mathbf{v}_{i,m}\in \mathbb{C}^{N_{t} \times 1}$ denotes the precoding vector for the $i$-th data stream $s_{i,m}$ of $\mathbf{s}_m$.
Then, the signal received by the $i$-th antenna of user $k$ is given by
\vspace{-6pt} 
\begin{equation}
\label{eqa:10}
\begin{aligned}
y_{i, k}= & \underbrace{\mathbf{h}_{i, k}^H \mathbf{v}_{i, k} s_{i, k}}_{\text {desired signal }}+\underbrace{\mathbf{h}_{i, k}^H \sum_{m=1, m \neq k}^K \sum_{j=1}^{N_r} \mathbf{v}_{j, m} s_{j, m}}_{\text {interference from other user's streams }}\\
&+\underbrace{\mathbf{h}_{i, k}^H \sum_{t=1, t \neq i}^{N_r} \mathbf{v}_{t, k} s_{t, k}}_{\text {interference from same user's streams }}+n_{i, k},
\end{aligned}
\end{equation}
where $n_{i, k}\sim \mathcal{CN}(0,\sigma_k^2)$ represents the noise on the $i$-th receive antenna.

The SE of the $i$-th antenna for user $k$ can be expressed as $R_{i,k}$ in (11) (see the bottom of next page). Based on this, the following optimization problem $\mathcal{P}_3$ can be formulated to obtain the precoding matrix that maximizes the system's WSR.

\begin{figure*}[b]
\vspace{-10pt} 
\centering
\hrule 
\vspace{-1pt} 
\begin{align}
R_{i,k}=\log_2 \left(1 + \frac{\left|\mathbf{h}_{i,k}^H \mathbf{v}_{i,k}\right|^2}{\sum_{m=1, m \neq k}^K \sum_{j=1}^{N_r}\left|\mathbf{h}_{i,k}^H \mathbf{v}_{j,m}\right|^2+ \sum_{t=1, t \neq i}^{N_r}\left|\mathbf{h}_{i,k}^H \mathbf{v}_{t,k}\right|^2 +\sigma_k^2}\right),\forall i, \forall k.
\end{align}
\end{figure*}
\vspace{-6pt} 
\begin{equation}
\label{eqa:12}
\begin{aligned}
\mathcal{P}_3: \max_{\left\{\mathbf{v}_{i,k}\right\}} & \sum_{k=1}^K \sum_{i=1}^{N_r}\alpha_k R_{i,k} \\
\text {s.t.} & \sum_{k=1}^K  \sum_{i=1}^{N_r}\operatorname{Tr}\left(\mathbf{v}_{i,k} \mathbf{v}_{i,k}^H\right) \leq P_t.
\end{aligned}
\end{equation}

In Fig. 3(c), each VN includes a single-antenna receiver. For ease of representation, the VN corresponding to the $i$-th antenna of user $k$ is denoted as $v_{k,i}$. The VN set $\mathit{V}$ can be denotes as $\mathit{V}=\left\{\mathit{v}_{1,1}, \mathit{v}_{1,2}, \ldots, \mathit{v}_{1,{N_r}}, \ldots,\mathit{v}_{K,{N_r}} \right\}$ and $|V|=KN_r$. Thus, the index of node $v_{k,i}$ in the set is $((k-1)N_r+i)$. The channel feature of the VN $v_{k,i}$ can be represented as $\mathbf{X}^{rc}_{((k-1)N_r+i,:)}=\mathbf{h}_{i,k}$, where $\mathbf{X}^{rc}\in \mathbb{C}^{KN_{r} \times N_{t}}$, i.e., $d_2=N_t$. The edge feature tensor $\mathcal{H}\in \mathbb{C}^{KN_{r} \times KN_{r} \times N_{t}}$ is given by
\begin{equation}
\mathcal{H}\!\!=\!\!\begin{Bmatrix}
\mathbf{0}      & \mathbf{h}_{2,1} & \cdots & \mathbf{h}_{N_r,1} & \cdots& \mathbf{h}_{N_r,K} \\
\mathbf{h}_{1,1}& \mathbf{0}       & \cdots & \mathbf{h}_{N_r,1} & \cdots& \mathbf{h}_{N_r,K} \\
\vdots & \vdots & \ddots & \vdots & \ddots & \vdots \\
\mathbf{h}_{1,1}& \mathbf{h}_{2,1} & \cdots & \mathbf{0}         & \cdots& \mathbf{h}_{N_r,K} \\
\mathbf{h}_{1,1}& \mathbf{h}_{2,1} & \cdots & \mathbf{h}_{N_r,1} & \cdots& \mathbf{0}
\end{Bmatrix}_{KN_{r} \times KN_{r}\times N_{t}},
\end{equation}
where $\mathbf{0} \in \mathbb{C}^{N_{t} \times 1}$ is a zero vector. \textit{Note:} $\mathcal{H}$ represents a three-dimensional tensor. According to (10), both the interference and desired signals received by the VN $v_{k,i}$ pass through the same channel to the receiver. Therefore, the features of the directed edges to $v_{k,i}$ are all characterized by the channel vector $\mathbf{h}_{i,k}$. The optimization variable for VN $v_{k,i}$ is $\mathbf{\Gamma}_{((k-1)N_r+i,:)}=\mathbf{v}_{i,k}$, where $\mathbf{\Gamma} \in \mathbb{C}^{KN_r\times \mathit{N_t}}$. Consequently, the SE for the $i$-th antenna of user $k$ can be redefined as (14), where $\mathcal{H}(v_{m,j},v_{k,i},:)$ corresponds to $\mathcal{H}_{((m-1)N_r+j,(k-1)N_r+i,:)}$, and $\mathcal{H}(v_{k,t},v_{k,i},:)$ to $\mathcal{H}_{((k-1)N_r+t,(k-1)N_r+i,:)}$. 

\begin{figure*}[b]
\vspace{-10pt} 
\centering
\hrule 
\vspace{-1pt} 
\begin{align}
\widetilde{R}_{i,k}=\log_2 \left(1 + \frac{\left|\mathbf{X}_{((k-1)N_r+i,:)}^{rcH} \mathbf{\Gamma}_{((k-1)N_r+i,:)}\right|^2}{\sum_{m=1, m \neq k}^K \sum_{j=1}^{N_r}\left|\mathcal{H}(v_{m,j},v_{k,i},:)^H \mathbf{\Gamma}_{((m-1)N_r+j,:)}\right|^2+ \sum_{t=1, t \neq i}^{N_r}\left|\mathcal{H}(v_{k,t},v_{k,i},:)^H \mathbf{\Gamma}_{((k-1)N_r+t,:)}\right|^2 +\sigma_k^2}\right).
\end{align}
\end{figure*}
From this, we can establish the following graph optimization problem $\mathcal{P}_4$:
\vspace{-6pt}
\begin{equation}
\label{eqa:15}
\begin{aligned}
\mathcal{P}_4: &\max_{\left\{\mathbf{\Gamma}\right\}} \quad g(\boldsymbol{\alpha},\mathbf{\Gamma},\mathbf{X}^{rc},\mathcal{H})= \sum_{k=1}^K \sum_{i=1}^{N_r}\alpha_k \widetilde{R}_{i,k} \\
&\text {s.t.} \quad C(\mathbf{\Gamma},\mathbf{X}^{rc},\mathcal{H})=\sum_{k=1}^K  \sum_{i=1}^{N_r}\operatorname{Tr}\left(\mathbf{v}_{i,k} \mathbf{v}_{i,k}^H\right) - P_t \leq 0.
\end{aligned}
\end{equation}
From (14) and (15), it can be observed that both the objective function $ g(\boldsymbol{\alpha},\mathbf{\Gamma},\mathbf{X}^{rc},\mathcal{H})$ and constraint function $C(\mathbf{\Gamma},\mathbf{X}^{rc},\mathcal{H})$ are composed of symmetric operations. Based on \textit{Proposition 1}, they also exhibit the permutation invariance property.

\subsection{The ICGNN for MU-MIMO System}
In a MU-MIMO system, for the VN $v_{k,i}$, the channel feature is $\mathbf{h}_{i,k}$, and the target feature is the precoding vector $\mathbf{v}_{i,k}$. Given that most neural networks only process real data, the feature update network $\mathrm{U}(\cdot)$ need to output both the real and imaginary parts separately, i.e., $(\Re(\mathbf{v}_{i,k})^T, \Im(\mathbf{v}_{i,k})^T)\in \mathbb{R}^{1\times 2N_t}$, where the output layer is composed of $2N_t$ neurons. Using $\mathbf{v}_{i,k}$ as the output for the $\mathrm{U}(\cdot)$ network  not only leads to a larger scale for the $\mathrm{M}(\cdot)$ and $\mathrm{U}(\cdot)$ networks, but also hinders the networks' convergence. To reduce the network size and enhance convergence speed, we find that $\mathcal{P}_3$ can be solved using the optimal solution structure for MISO beamforming \cite{ref27},
\vspace{-15pt}
\begin{equation}
\label{eqa:16}
\begin{aligned}
\mathbf{v}_{i,k}=\sqrt{p_{k,i}} \frac{\left(\mathbf{I}_{N_t}+\sum_{m=1}^K \sum_{j=1}^{N_r}\frac{\lambda_{m,j}}{\sigma_m^2} \mathbf{h}_{j,m} \mathbf{h}_{j,m}^H\right)^{-1} \mathbf{h}_{i,k}}{\left\|\left(\mathbf{I}_{N_t}+\sum_{m=1}^K \sum_{j=1}^{N_r} \frac{\lambda_{m,j}}{\sigma_m^2} \mathbf{h}_{j,m} \mathbf{h}_{j,m}^H\right)^{-1} \mathbf{h}_{i,k}\right\|},\\
\forall i, \forall k,
\end{aligned}
\end{equation}
where $p_{k,i}\geq 0$ and $\lambda_{k,i}\geq 0$ represent the downlink and equivalent uplink power allocation factor and for the $i$-th antenna of user $k$, respectively. They satisfy $\sum_{k=1}^K \sum_{i=1}^{N_r}p_{k,i}=\sum_{k=1}^K \sum_{i=1}^{N_r}\lambda_{k,i}=P_t$. Thus, the precoding vector $\mathbf{v}_{i,k}$ can be calculated using the power allocation factors $\{p_{k,i}, \lambda_{k,i}, \forall i, \forall k\}$ with (16), allowing the feature update network $\mathrm{U}(\cdot)$ to only output $\{p_{k,i}, \lambda_{k,i}, \forall i, \forall k\}$, significantly reducing the size of the output layer from $2N_t$ to $2$. Therefore, in the ICGNN, the feature of $v_{k,i}$ is defined as $\mathbf{X}_{((k-1)N_r+i,:)}=(\Re(\mathbf{h}_{i,k})^T, \Im(\mathbf{h}_{i,k})^T, p_{k,i}, \lambda_{k,i})^T$, where $\mathbf{X}\in \mathbb{R}^{KN_r\times (2N_t+2)}$. Here, the initial values for $p_{k,i}$ and $\lambda_{k,i}$ are set to equal power distribution, i.e., $\{p_{k,i}= \lambda_{k,i}=\frac{P_t}{KN_r}, \forall i, \forall k\}$. The feature of the edge from $v_{m,j}$ to $v_{k,i}$ can be defined as $\mathcal{H}(v_{m,j},v_{k,i},:)=(\Re(\mathbf{h}_{i,k})^T, \Im(\mathbf{h}_{i,k})^T)^T$, where $\mathcal{H}\in \mathbb{R}^{KN_r\times KN_r\times2N_t}$. 
The neuron units for each layer of $\mathrm{M}(\cdot)$ and $\mathrm{U}(\cdot)$ networks are set as $\{4N_t+2, 128, 256, 64\}$ and $\{2N_t+66, 128, 32, 2\}$, respectively, with a batch normalization layer and Tanh activation function added after the hidden layers. The output layers of $\mathrm{M}(\cdot)$ and $\mathrm{U}(\cdot)$ use Tanh and Sigmoid activation functions, respectively, with the Sigmoid ensuring the output power factors are positive.
Finally, we need to scale the power factors output by $\mathrm{U}(\cdot)$ to ensure they meet the power constraints through the following transformation,
\vspace{-5pt}
\begin{equation}
\label{eqa:17}
\begin{aligned}
p_{k,i}(\Theta)= \frac{p_{k,i}(\Theta)P_t}{\sum_{k=1}^K \sum_{i=1}^{N_r}p_{k,i}(\Theta)},\\
\lambda_{k,i}(\Theta)=\frac{\lambda_{k,i}(\Theta)P_t}{\sum_{k=1}^K \sum_{i=1}^{N_r}\lambda_{k,i}(\Theta)},
\end{aligned}
\vspace{-5pt}
\end{equation}

where $\Theta$ represents the network weights of the ICGNN.

The network is trained using an unsupervised learning approach, with the loss function defined as follows:
\vspace{-6pt}
\begin{subequations}
\label{eqa:18}
\renewcommand{\theequation}{\theparentequation\alph{equation}}  
\begin{align}
\mathcal{L}_1(\Theta) &=-\mathbb{E}\left(\sum_{k=1}^K \alpha_k\log_2 \operatorname{det}\left(\mathbf{I} + \mathrm{SINR}_k\right)\right), \\ 
\mathrm{SINR}_k &= \mathbf{H}_k^H \mathbf{V}_k(\Theta) \mathbf{V}_k^H(\Theta) \mathbf{H}_k \cdot \nonumber \\
&\left(\sum_{m \neq k}^K \mathbf{H}_k^H \mathbf{V}_m(\Theta) \mathbf{V}_m^H(\Theta) \mathbf{H}_k + \sigma_k^2 \mathbf{I}\right)^{-1},
\end{align}
\end{subequations}
where $\mathbf{V}_k(\Theta)=[\mathbf{v}_{1,k}(\Theta), \mathbf{v}_{2,k}(\Theta),\ldots, \mathbf{v}_{N_r,k}(\Theta)]$ represents the precoding matrix recovered from $p_{k,i}(\Theta)$ and $\lambda_{k,i}(\Theta)$ using (16), and the expectation $\mathbb{E}(\cdot)$ is taken over all the channel realizations.  Due to the fact that neural networks do not support complex-valued operations, it is necessary to map complex-valued operations to the real-valued field. We adopt the decomposition from \cite{ref28} as follows: 
$\mathbf{V}_k(\Theta)=\left[\begin{array}{cc}\Re(\mathbf{V}_k(\Theta)) & -\Im(\mathbf{V}_k(\Theta)) \\ \Im(\mathbf{V}_k(\Theta)) & \Re(\mathbf{V}_k(\Theta))\end{array}\right]$, $\mathbf{H}_k=\left[\begin{array}{cc}\Re(\mathbf{H}_k) & -\Im(\mathbf{H}_k) \\ \Im(\mathbf{H}_k) & \Re(\mathbf{H}_k)\end{array}\right]$.

\subsection{The Matrix-Inverse-Free Precoding Vector Recovery Scheme}
The high-dimensional matrix inversion operation is a significant barrier to the practical application of the WMMSE algorithm. The precoding vector given in (16) involves an $N_t$-dimensional matrix inversion, which results in high computational complexity in large-scale MIMO scenarios. We now present a low-complexity precoding vector recovery scheme that does not involve matrix inversion.
Equation (16) is valid based on the fact that $\mathbf{A}=\mathbf{I}_{N_t}+\sum_{m=1}^K \sum_{j=1}^{N_r}\frac{\lambda_{m,j}}{\sigma_m^2} \mathbf{h}_{j,m} \mathbf{h}_{j,m}^H$ is invertible because $\mathbf{A}$ is strongly diagonally dominant, meaning $\mathbf{A}\mathbf{v}_{i,k}= \frac{\sqrt{p_{k,i}}\mathbf{h}_{i,k}}{\left\|\mathbf{A}^{-1} \mathbf{h}_{i,k}\right\|}$ has a unique solution. Since $\frac{\sqrt{p_{k,i}}}{\left\|\mathbf{A}^{-1} \mathbf{h}_{i,k}\right\|}$ is a scalar, let's denote $q_{k,i}=\frac{\sqrt{p_{k,i}}}{\left\|\mathbf{A}^{-1} \mathbf{h}_{i,k}\right\|}$. We find that the coefficient matrix $\mathbf{A}$ is conjugate symmetric, making it suitable for solving using the CG method\cite{ref29}.

The CG method is an iterative solving process. First, initialize $\mathbf{v}_{i,k}^{(0)}=\mathbf{h}_{i,k}$ using the maximal ratio transmission (MRT) method\cite{ref2}.
Assuming the existence of a conjugate set $\{\mathbf{p}_0, \mathbf{p}_1, \dots, \mathbf{p}_{N_t-1}\}$ with respect to $\mathbf{A}$, where $\mathbf{p}_i^{H}\mathbf{A}\mathbf{p}_j=0, \forall i\neq j $, the solution $\mathbf{v}_{i,k}^{(*)}$ must be found within at most $N_t$ steps by iterating as follows.
\begin{equation}
\label{eqa:19}
\begin{aligned}
\mathbf{v}_{i,k}^{(t+1)}=\mathbf{v}_{i,k}^{(t)}+\delta_t\mathbf{p}_t, \forall t=0,1,\dots,N_t-1,
\end{aligned}
\end{equation}
where $\delta_t=-\frac{\mathbf{r}_{i,k}^{(t)^H}\mathbf{p}_t}{\mathbf{p}_t^{H}\mathbf{A}\mathbf{p}_t}$ is the step size, determined through a step-length selection algorithm \cite{ref29}, where $\mathbf{r}_{i,k}^{(t)}=\mathbf{A}\mathbf{v}_{i,k}^{(t)}-q_{k,i}\mathbf{h}_{i,k}=\mathbf{r}_{i,k}^{(t-1)}+\delta_{t-1}\mathbf{A}\mathbf{p}_{t-1}$. 
There is a proof provided below.

\textit{Proof}: The conjugate set $\{\mathbf{p}_0, \mathbf{p}_1, \dots, \mathbf{p}_{N_t-1}\}$ is a set of linearly independent vectors. By the nature of N-dimensional linear space, we know that $\mathbf{v}_{i,k}^{(*)}-\mathbf{v}_{i,k}^{(0)}$ can be uniquely represented by $\{\mathbf{p}_0, \mathbf{p}_1, \dots, \mathbf{p}_{N_t-1}\}$, i.e., $\mathbf{v}_{i,k}^{(*)}-\mathbf{v}_{i,k}^{(0)}=\delta_0\mathbf{p}_{0}+\delta_1\mathbf{p}_{1}+\dots+\delta_{N_t-1}\mathbf{p}_{N_t-1}.$
It means that by selecting appropriate scalars $\{\delta_t, \forall t\}$ and iterating $N$ times from the starting point $\mathbf{v}_{i,k}^{(0)}$, it can converge to the solution $\mathbf{v}_{i,k}^{(*)}$, i.e., $\mathbf{v}_{i,k}^{(*)}=\mathbf{v}_{i,k}^{(0)}+\sum_{t=0}^{N_t-1}\delta_t\mathbf{p}_{t}$.

Each conjugate vector $\mathbf{p}_t$ in the CG method is chosen to be a linear combination of the negative gradient $-\mathbf{r}_{i,k}^{(t)}$ and the previous conjugate vector $\mathbf{p}_{t-1}$, and the initial value is set as $\mathbf{p}_0=-\mathbf{r}_{i,k}^{(0)}$. This combination can be formally represented as:
\begin{equation}
\label{eqa:20}
\begin{aligned}
\mathbf{p}_t = -\mathbf{r}_{i,k}^{(t)}+\beta_t\mathbf{p}_{t-1}.
\end{aligned}
\end{equation}
where the scalar $\beta_t$ is to guarantee the requirement that $\mathbf{p}_t$ and $\mathbf{p}_{t-1}$ must be conjugate with respect to $\mathbf{A}$, i.e., $\mathbf{p}_{t-1}^H\mathbf{A}\mathbf{p}_t=0$. According to the \textit{Expanding Subspace Minimization theorem} [29, Theorem 5.2], i.e., $\mathbf{r}_{i,k}^{(t)^H}\mathbf{p}_{j}=0, \forall j=0,1,\dots,t-1$, $\beta_t$ can be denoted as $\beta_{t} = \frac{\mathbf{r}_{i,k}^{(t)^H}\mathbf{r}_{i,k}^{(t)}}{\mathbf{r}_{i,k}^{(t-1)^H}\mathbf{r}_{i,k}^{(t-1)}}$. An outline of the proposed matrix-inverse-free method is given in \textbf{Algorithm 1}. The number of iterations $T$ can be selected based on the trade-off between complexity and performance in practical applications.  Moreover, $T\leq N_t$ always holds, based on the following theorem.

\begin{algorithm}
\caption{The matrix-inverse-free recovery method.}\label{alg:alg1}
\begin{algorithmic}
\STATE \hspace{-0.2cm}\textbf{Initialize}:\kern1.0ex$\mathbf{v}_{i,k}^{(0)}=\mathbf{h}_{i,k}$, $\mathbf{r}_{i,k}^{(0)}=\mathbf{A}\mathbf{v}_{i,k}^{(0)}-q_{k,i}\mathbf{h}_{i,k}$, $\mathbf{p}_{0}=-\mathbf{r}_{i,k}^{(0)}$, and iteration count $T$. 
\STATE \hspace{-0.2cm}\textbf{for} $ t = 0,1,...,T-1 $\textbf{ do }
\STATE \hspace{0.2cm}\textbf{calculate step size}\kern1.0ex: $\delta_t=-\frac{\mathbf{r}_{i,k}^{(t)H}\mathbf{p}_t}{\mathbf{p}_t^{H}\mathbf{A}\mathbf{p}_t}$.
\STATE \hspace{0.2cm}\textbf{update}\kern1.0ex$\mathbf{v}_{i,k}$\kern1.0ex: $\mathbf{v}_{i,k}^{(t+1)}=\mathbf{v}_{i,k}^{(t)}+\delta_t\mathbf{p}_t$.
\STATE \hspace{0.2cm}\textbf{update gradient}\kern1.0ex$\mathbf{r}_{i,k}$\kern1.0ex: $\mathbf{r}_{i,k}^{(t+1)}=\mathbf{r}_{i,k}^{(t)}+\delta_{t}\mathbf{A}\mathbf{p}_{t}$.
\STATE \hspace{0.2cm}\textbf{calculate}\kern1.0ex$\beta$:\kern1.0ex $\beta_{t+1} = \frac{\mathbf{r}_{i,k}^{(t+1)^H}\mathbf{r}_{i,k}^{(t+1)}}{\mathbf{r}_{i,k}^{(t)^H}\mathbf{r}_{i,k}^{(t)}}$.
\STATE \hspace{0.2cm}\textbf{update}\kern1.0ex$\mathbf{p}$:\kern1.0ex $\mathbf{p}_{t+1} = -\mathbf{r}_{i,k}^{(t+1)}+\beta_{t+1}\mathbf{p}_{t}$.
\STATE \hspace{-0.2cm}\textbf{Output}:\kern1.0ex $\mathbf{v}_{i,k}^{(T)}$.
\STATE \hspace{-0.2cm}\textbf{Power constraint}:\kern1.0ex $\mathbf{v}_{i,k}^{(T)}=\mathbf{v}_{i,k}^{(T)}\sqrt{\frac{P_t}{\sum_{k=1}^K  \sum_{i=1}^{N_r}\operatorname{Tr}\left(\mathbf{v}_{i,k}^{(T)} \mathbf{v}_{i,k}^{(T)^H}\right)}}$
\end{algorithmic}
\label{alg1}
\vspace{-4pt}
\end{algorithm}
\textit{Theorem 1 (The CG method converges in $r$ iterations, $r\leq N_t$): If $\mathbf{A}$ has only $r$ distinct eigenvalues, then the CG iteration will terminate at the solution in at most $r$ iterations, i.e., $\left\|\mathbf{v}_{i,k}^{(r)}-\mathbf{v}_{i,k}^{(*)}\right\|_{\mathbf{A}}^2 =c_{\rm rate}= 0$, where $\left\|\mathbf{v}\right\|_{\mathbf{A}}^2=\mathbf{v}^H\mathbf{A}\mathbf{v}$ and $c_{\rm rate}$ is a nonnegative constant. Additionally, if there are $N_t$ distinct eigenvalues distributed among $r$ distinct clusters, the CG method can also approximately converge to the optimal solution $\mathbf{v}_{i,k}^{(*)}$ within $r$ iterations, i.e., $\frac{\left\|\mathbf{v}_{i,k}^{(r)}-\mathbf{v}_{i,k}^{(*)}\right\|_{\mathbf{A}}^2}{\left\|\mathbf{v}_{i,k}^{(0)}-\mathbf{v}_{i,k}^{(*)}\right\|_{\mathbf{A}}^2}=c_{\rm rate}\approx 0$.}

\textit{Proof}: 
The convergence rate of the CG method can be quantified by estimating the nonnegative scalar quantity $c_{\rm rate} =\min_{P_d} \max_{1 \leq i \leq N_t} \left[1 + \lambda_i P_d(\lambda_i)\right]^2$, where $P_d$ denotes a polynomial of degree $d$ and $\{\lambda_i\}_{i=1}^{N_t}$ are the eigenvalues of $\mathbf{A}$ \cite{ref29}. We aim to find a polynomial $P_d$ that minimizes $c_{\rm rate}$ as much as possible, which implies convergence of the CG method.

If $\mathbf{A}$ has only $r$ distinct eigenvalues $\{\bar{\lambda}_1,\ldots, \bar{\lambda}_r\}$, then a polynomial of degree $r-1$ can be found, i.e., $P_{r-1}(\lambda) = \frac{Q_r(\lambda) - 1}{\lambda}$, where $Q_r(\lambda)$ can be defined as follows:
\begin{equation}
\label{eqa:cg}
\begin{aligned}
Q_r(\lambda)=\frac{(-1)^r}{\prod_{i=1}^{r}\bar{\lambda}_i}\prod_{i=1}^{r}\left(\lambda-\bar{\lambda}_i \right).
\end{aligned}
\end{equation}
Substituting $P_{r-1}(\lambda)$ into $c_{\rm rate}$ yields $c_{\rm rate} = 0$, which indicates that convergence can be achieved in at most $r$ iterations.
Similarly, if $\mathbf{A}$ has $N_t$ distinct eigenvalues distributed across $r$ clusters, we can construct a polynomial $P_{r-1}(\lambda)$ such that the roots of $1 + \lambda P_{r-1}(\lambda)$ lie within each cluster. Substituting this into $c_{\rm rate}$ yields $c_{\rm rate} < \epsilon$, where $\epsilon$ is a small positive constant\cite{ref30}. This indicates that the CG method can approximately converge within $r$ iterations. The above analysis completes the proof of the convergence rate for the CG method.

This implies that convergence to the optimal solution in $N_t$ iterations occurs only under the worst-case scenario, where the eigenvalues of the coefficient matrix $\mathbf{A}$ are widely spaced. In \textbf{Algorithm 1}, the primary computational effort at each iteration involves the calculation of the matrix-vector product $\mathbf{A}\mathbf{p}_t$. The algorithm's complexity is $\mathcal{O}(T N_t^2)$, where $T \leq N_t$, which is less than the $\mathcal{O}(N_t^3)$ complexity of the precoding recovery scheme presented in equation (16).

\begin{figure*}[!t]
\vspace{-10pt}
\centering
\includegraphics[scale=0.6]{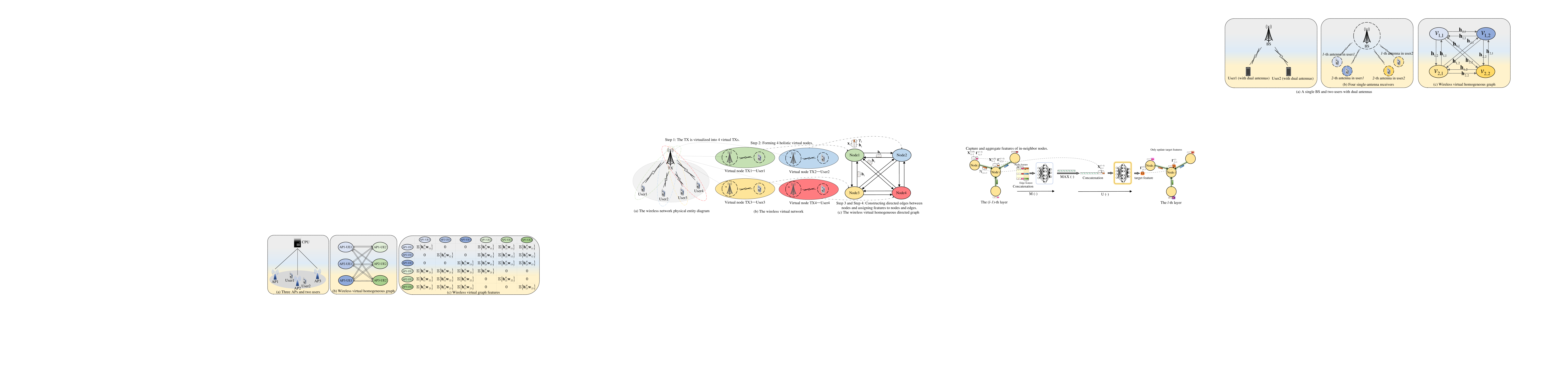}
\vspace{-10pt}
\caption{Wireless virtual graph of a cell-free MIMO system.}
\label{fig:4}
\vspace{-15pt}
\end{figure*}

\section{A Scalable Power Allocation Scheme in Cell-Free Scenario}
\subsection{System Model and Problem Formulation}
This section discusses a cell-free massive MIMO system with $L$ APs, each equipped with $N_t$ antennas, serving $K$ single-antenna users. The channel between user $k$ and AP $l$ is denoted by $\mathbf{h}_{kl}\in \mathbb{C}^{N_t\times 1}$. 
In general, downlink data transmission in such systems can be implemented in two modes: centralized or distributed operation. In this section, we consider a dynamic cell-free scenario in which the number of serving APs and served users varies frequently due to user mobility. This requires each AP to rapidly determine beamforming and power allocation strategies within a short time frame. However, the high fronthaul link overhead associated with centralized operation, such as the collection of uplink pilot signals from all APs\cite{add1}, makes it less suitable for such scenarios. Consequently, a distributed operation is adopted. This system configuration yields a two-stage process: each AP first performs beamforming based on locally estimated channels, followed by a network-wide power allocation decision made at the central processing unit (CPU).

Under the assumption that all users are simultaneously served by all $L$ APs, the received signal for user $k$ can be represented as
\vspace{-6pt}
\begin{equation}
\label{eqa:21}
\begin{aligned}
y_k=\sum_{l=1}^L \mathbf{h}_{kl}^H \sum_{i=1}^K \sqrt{p_{i l}} \mathbf{w}_{i l} s_i+n_k,
\end{aligned}
\vspace{-6pt}
\end{equation}
where $s_i \sim \mathcal{CN}(0,1)$ represents the desired received signal for user $i$, $p_{i l}\geq 0$ denotes the power allocated by AP $l$ to user $i$, $\mathbf{w}_{il}$ is the normalized precoding vector, and $n_k\sim \mathcal{CN}(0,\sigma^2)$ represents the receiver noise. We focus solely on the network-wide power allocation, while $\mathbf{w}_{kl}$ is chosen to be the local minimum mean square error (L-MMSE) precoding scheme defined as \cite{ref31}
\vspace{-6pt}
\begin{equation}
\label{eqa:22}
\begin{aligned}
\mathbf{w}_{kl}=\frac{\left(\sum_{i=1}^K \rho_i \mathbf{h}_{i l} \mathbf{h}_{i l}^H+\sigma^2 \mathbf{I}_{N_t}\right)^{-1} \rho_k \mathbf{h}_{k l}}{\|\left(\sum_{i=1}^K \rho_i \mathbf{h}_{i l} \mathbf{h}_{i l}^H+\sigma^2 \mathbf{I}_{N_t}\right)^{-1} \rho_k \mathbf{h}_{k l}\|},
\end{aligned}
\vspace{-6pt}
\end{equation}
where $\rho_i$ is the uplink transmit power of user $i$.


\textit{Lemma 1 [32, Lem. 1]:} \textit{When the precoding vector that satisfies the condition $\mathbb{E}\left\{\mathbf{h}_{k l}^H \mathbf{w}_{k l}\right\}\geq0$, the lower bound of the downlink SE for user $k$ can be expressed as}
\vspace{-6pt}
\begin{equation}
\label{eqa:23}
\begin{aligned}
\mathrm{SE}_k=\frac{\tau_d}{\tau_c} \log _2\left(1+\mathrm{SINR}_k\right),
\end{aligned}
\vspace{-6pt}
\end{equation}
\textit{where}
\vspace{-2pt}
\begin{equation}
\label{eqa:24}
\begin{aligned}
\operatorname{SINR}_k=\frac{\left(\mathbf{a}_k^T \boldsymbol{\mu}_k\right)^2}{\sum_{i=1}^K \boldsymbol{\mu}_i^T \mathbf{B}_{k i} \boldsymbol{\mu}_i-\left(\mathbf{a}_k^T \boldsymbol{\mu}_k\right)^2+\sigma^2},
\end{aligned}
\end{equation}
\vspace{-2pt}
\textit{represents the effective SINR, $\tau_c$ and $\tau_d$ respectively denote the duration of the coherence block and the duration within it used for downlink data transmission,}
\vspace{-3pt}
\begin{equation}
\label{eqa:25}
\begin{aligned}
& \boldsymbol{\mu}_k=\left[\mu_{k 1}, \ldots, \mu_{k L}\right]^T \in \mathbb{R}^{L \times 1}, \mu_{k l}=\sqrt{p_{k l}}, \\
& \mathbf{a}_k=\left[a_{k 1}, \ldots, a_{k L}\right]^T \in \mathbb{R}^{L \times 1}, a_{k l}=|\mathbb{E}\left\{\mathbf{h}_{k l}^H \mathbf{w}_{k l}\right\}|, \\
& \mathbf{B}_{k i} \in \mathbb{R}^{L \times L}, b_{k i}^{l m}=\Re\left(\mathbb{E}\left\{\mathbf{h}_{k l}^H \mathbf{w}_{i l} \mathbf{w}_{i m}^H \mathbf{h}_{k m}\right\}\right),
\end{aligned}
\end{equation}
\vspace{-1pt}
\textit{and $b_{k i}^{l m}$ refers to element $(l,m)$ in matrix $\mathbf{B}_{k i}$}.

Clearly, the L-MMSE precoding scheme satisfies $\mathbb{E}\left\{\mathbf{h}_{k l}^H \mathbf{w}_{k l}\right\}\geq0$. According to \textit{Lemma 1}, we can formulate the following WSR maximization power allocation problem,
\begin{equation}
\label{eqa:26}
\begin{aligned}
\mathcal{P}_5: \underset{\left\{\mu_{k l}: \forall k, l\right\}}{\operatorname{max}} & \frac{\tau_d}{\tau_c} \sum_{k=1}^K \alpha_k\log _2\left(1+\mathrm{SINR}_k\right) \\
\text { s.t. } & \sum_{k=1}^K \mu_{k l}^2 \leq P_{t}, \quad \forall l=1, \ldots, L ,
\end{aligned}
\end{equation}
where $P_{t}$ denotes transmit power for each AP. This problem is non-convex and can be addressed using the WMMSE algorithm combined with the alternating direction method of multipliers (ADMM) algorithm to obtain a local optimal solution [33, Alg. 1]. 
This method will also serve as a benchmark in the subsequent analysis.

\subsection{The ICGNN for Cell-Free MIMO Power Allocation}
We take three APs and two users as an example, depicted in Fig. 4(a), to construct a WVG for a cell-free network, as shown in Fig. 4(b). In Fig. 4, each AP is abstracted into the same number of virtual AP nodes as the number of users it serves. Unlike cellular networks, there is no interference between APs that simultaneously serve the same user $k$, hence, there are no directed edges between nodes ``AP$l$-UE$k$, $\forall l$''. 

We denote each VN ``AP$l$-UE$k$, $\forall l, \forall k$'' as $v_{kl}$, with the VN set $V$ defined as $V=\{v_{11}, v_{12},\ldots, v_{1L},\ldots, v_{KL}\}$, where $|V|=KL$. The index of $v_{kl}$ is $((k-1)L+l)$. We define the channel feature carried by $v_{kl}$ as $a_{kl}=\mathbb{E}\left\{\mathbf{h}_{k l}^H \mathbf{w}_{k l}\right\}\in \mathbb{R}$, representing the effective signal channel, shown on the diagonal in Fig. 4(c). The target feature of $v_{kl}$ is the power allocation coefficient $p_{kl}$, which satisfies $\sum_{k=1}^K p_{kl}=P_t, \forall l$. During network training, the target feature is initialized to $p_{kl}=\frac{1}{K}, \forall l, \forall k$, i.e., equal power distribution. Therefore, the feature of the VN $v_{kl}$ can be represented as $\mathbf{X}_{((k-1)L+l,:)}=(a_{kl}, p_{kl})^T$, where $\mathbf{X}\in \mathbb{R}^{KL\times2}$. Next, we define the effective interference channel as $I_{kl}^m=\mathbb{E}\left\{\mathbf{h}_{m l}^H \mathbf{w}_{k l}\right\}\in \mathbb{R}, \forall m\neq k$, which represents the interference to user $m$ from signals sent by AP $l$ to user $k$. The feature of the directed edge from $v_{kl}$ to $v_{mj}$ can be denoted as $\mathbf{H}(v_{kl},v_{mj})=\mathbf{H}_{((k-1)L+l, (m-1)L+j)}=I_{kl}^m \in \mathbb{R}$, where $\mathbf{H}\in \mathbb{R}^{KL\times KL}$, as illustrated on the non-diagonal lines in Fig. 4(c). 
We define the effective signal channel and effective interference channel as the feature of VNs and edges, respectively. On one hand, this approach allows the dimensions of graph's feature to be independent of the number of AP antennas, enabling the ICGNN to adapt to scenarios with variable antenna counts. On the other hand, this significantly reduces the feature dimensions, which helps in scaling down the size of the ICGNN network.
Furthermore, similar to Section III, our feature design ensures that changes in the number of users and APs only affect the size of the WVG, not the structure of the ICGNN. This means that the ICGNN can adapt to scenarios with variable numbers of users and APs.

The configuration of neuron units in the $\mathrm{M}(\cdot)$ and $\mathrm{U}(\cdot)$ networks is set as $\{3, 32, 128, 64\}$ and $\{66, 128, 32, 1\}$, respectively, with each layer's composition being identical to what is described in Section III C.
The power factors output by $\mathrm{U}(\cdot)$ need to be scaled to meet the power constraints, as shown below.

\vspace{-5pt}
\begin{equation}
\label{eqa:27}
\begin{aligned}
p_{kl}(\Theta)= \frac{p_{kl}(\Theta)P_t}{\sum_{k=1}^K p_{kl}(\Theta)},\forall l,
\end{aligned}
\end{equation}
where $\Theta$ also denotes the network weights of the ICGNN.

The network is trained through unsupervised learning methods, and the loss function is specified as follows:
\begin{subequations}
\label{eqa:28}
\renewcommand{\theequation}{\theparentequation\alph{equation}}  
\begin{align}
\mathcal{L}_2(\Theta) &=-\mathbb{E}\left(\frac{\tau_d}{\tau_c}\sum_{k=1}^K \alpha_k\log_2 \left(1 + \operatorname{SINR}_k\right)\right), \\ 
\operatorname{SINR}_k&=\frac{\left(\mathbf{a}_k^T \boldsymbol{\mu}_k(\Theta)\right)^2}{\sum_{i=1}^K \boldsymbol{\mu}_i^T(\Theta) \mathbf{B}_{k i} \boldsymbol{\mu}_i(\Theta)-\left(\mathbf{a}_k^T \boldsymbol{\mu}_k(\Theta)\right)^2+\sigma^2},
\end{align}
\end{subequations}
where $\boldsymbol{\mu}_k(\Theta)=[\sqrt{p_{k1}(\Theta)}, \sqrt{p_{k2}(\Theta)}, \ldots, \sqrt{p_{kL}(\Theta)}], \forall k$. 

\section{simulation results}
In this section, we present simulation results to validate the effectiveness of the proposed wireless homogeneous virtual graph construction method and the ICGNN in both cellular and cell-free network environments. Initially, we show the optimal performance of the ICGNN network trained across different configuration scenarios, followed by further tests on the generalization capabilities of the ICGNN in tables. For the simulations, a two-layer ICGNN architecture was implemented using Python 3.8 and Pytorch 11.2 on a GeForce GTX 3080Ti. During network training, the Adam optimizer \cite{ref34} was employed with a learning rate of 0.001 and a batch size of 100. Without loss of generality, each user was treated equally, and the user priority factor $\alpha_k=1, \forall k=1, \ldots, K$. All results are averaged over $10^4$ channel realizations.

\subsection{MU-MIMO Precoding for Cellular Networks}
We consider a cellular network where the BS is equipped with $N_t=16$ antennas, and users equipped with multiple antennas are randomly distributed within a 500-meter radius from the BS. The transmit power of the BS is $P_t = 10$ dBm. The total system bandwidth is 20 MHz, and the receiver noise power spectral density is $-174$ dBm/Hz. The channel between the BS and users is generated based on Rayleigh fading with pathloss. The pathloss is set at $-30.5 - 36.7\log_{10}(d)$ [dB] \cite{add1}, where $d$ represents the distance in meters between the BS and the users. This pathloss model matches well with the 3GPP Urban Microcell model\cite{3GPP}. 
Then, we consider the following benchmarks for comparison.
\begin{itemize}
\item{\textit{WMMSE}\cite{ref5}: The WMMSE performance metrics, used for comparison in the following subsections, are based on results after 100 iterations, with initial precoding matrices generated using the MRT technique\cite{ref2}.}
\item{\textit{MRT}\cite{ref2}: A heuristic precoding scheme that features low computational complexity and is suitable for low signal to noise ratio (SNR) scenarios.}
\item{\textit{WCGCN}\cite{ref19}: A GNN based beamforming approach originally designed for ad hoc networks, modified for use in MU-MIMO scenarios.}
\end{itemize}

\begin{figure}[!t]
\centering
\includegraphics[scale=0.55]{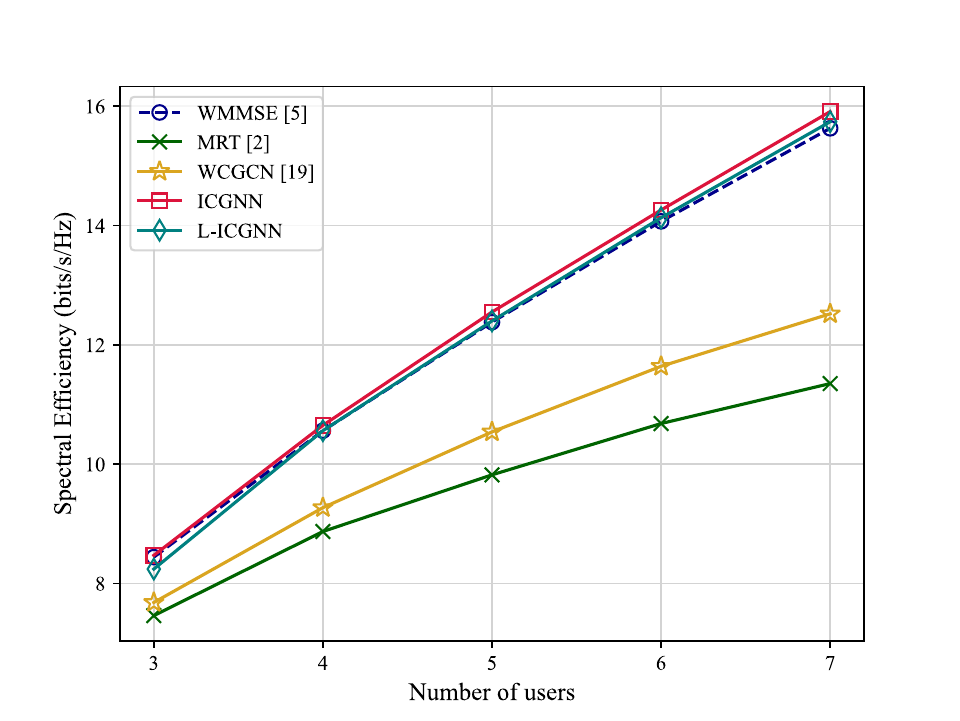}
\caption{SE versus different numbers of users ($N_r=2$).}
\label{fig:5}
\vspace{-8pt}
\end{figure}

\begin{table}[t]  
\begin{center}   
\caption{SE (bits/s/Hz) of A Model Trained on $K=5$ Users and Its Generalization to Different Numbers of Users.}  
\label{table:1} 
\begin{tabular}{|m{0.8cm}<{\centering}|m{1.1cm}<{\centering}|m{1.1cm}<{\centering}|m{1.8cm}<{\centering}|m{1.8cm}<{\centering}|}   
\hline   \textbf{No. of UEs} & \textbf{WMMSE}& \textbf{MRT}& \textbf{ICGNN}& \textbf{L-ICGNN} \\   
\hline   $3$ & $8.48$ & $7.51$ & $8.11(95.6\%)$ & $8.32(98.1\%)$ \\ 
\hline   $4$ & $10.57$ & $8.87$ & $10.52(99.5\%)$ & $10.53(99.6\%)$  \\  
\hline   $5$ & $12.38$ & $9.82$ & $12.55(101.4\%)$ & $12.40(100.2\%)$  \\   
\hline   $6$ & $14.07$ & $10.68$ & $14.31(101.7\%)$ & $14.11(100.3\%)$ \\     
\hline   $7$ & $15.63$ & $11.35$ & $15.90(101.7\%)$ & $15.67(100.3\%)$  \\  
\hline   
\end{tabular}   
\end{center}   
\end{table}

\begin{figure}[!t]
\centering
\includegraphics[scale=0.55]{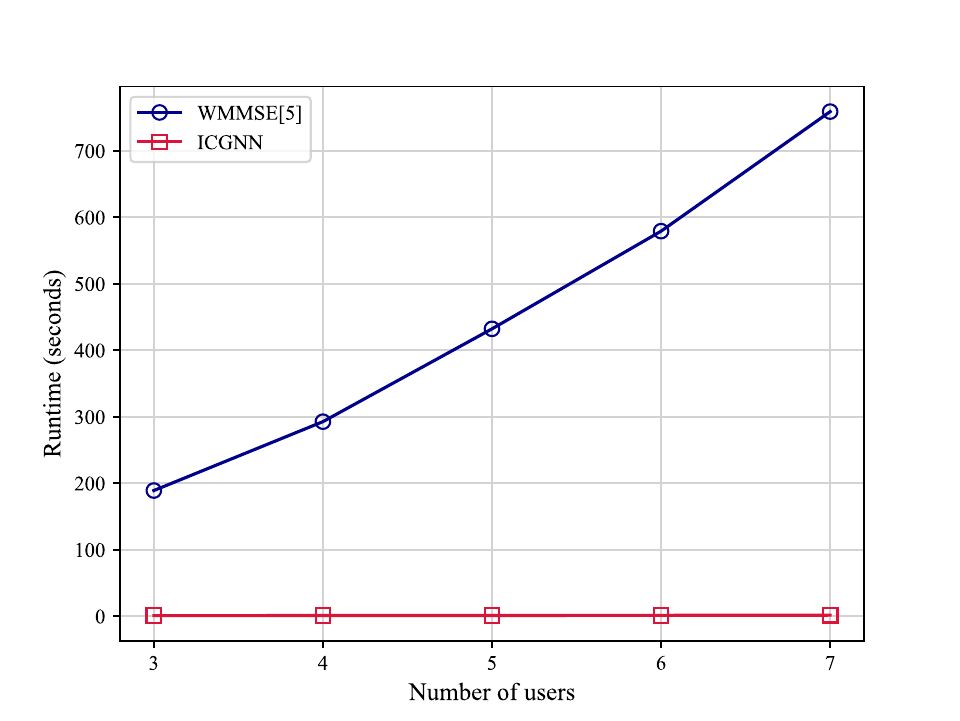}
\caption{Comparison of runtime between WMMSE and ICGNN.}
\label{fig:6}
\vspace{-18pt}
\end{figure}

\begin{figure}[!t]
\centering
\includegraphics[scale=0.55]{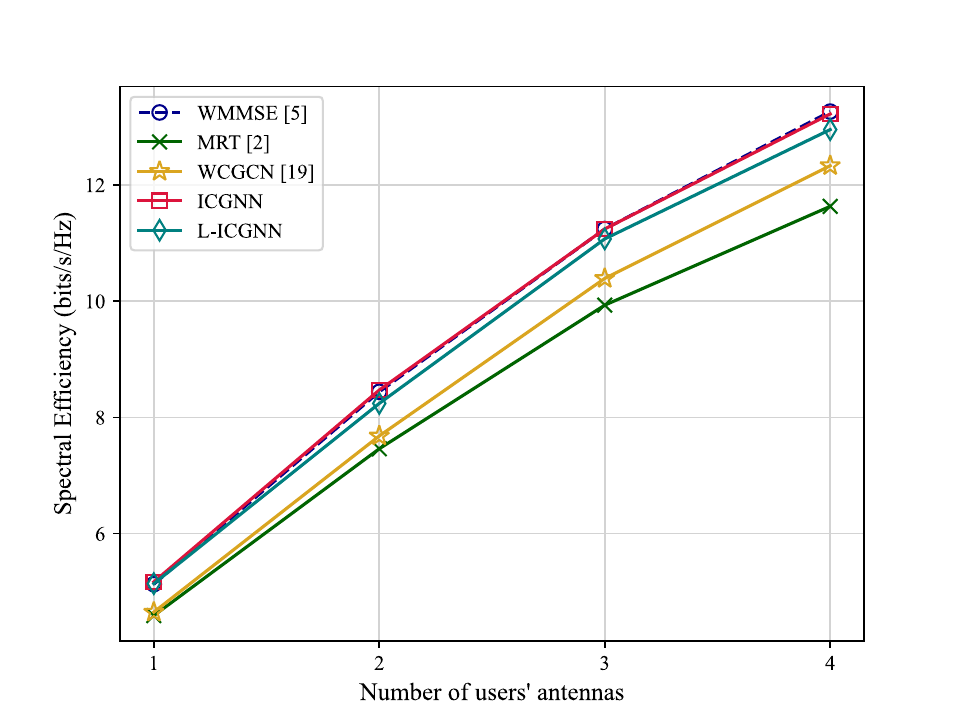}
\caption{SE versus different numbers of users' antennas ($K=3$).}
\label{fig:7}
\end{figure}

\begin{figure}[!t]
\centering
\includegraphics[scale=0.55]{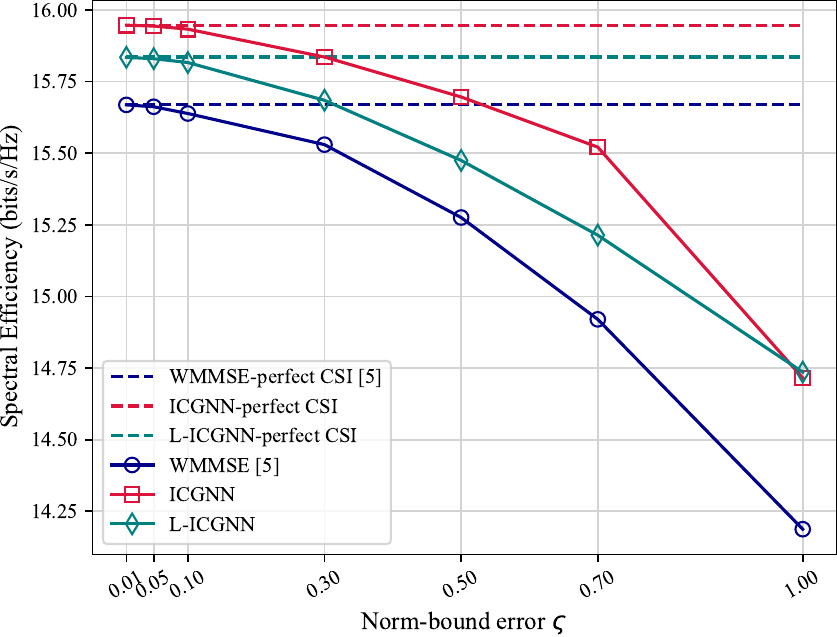}
\caption{Performance comparison under different norm-bounded errors $\varsigma$ ($K=7$, $N_r=2$).}
\end{figure}

\textit{1)} \textit{Generalization to Different Numbers of Users:}
In this subsection, we demonstrate the performance of comparative schemes with different user counts, each having $N_r=2$ antennas\footnote{To ensure sufficient spatial degrees of freedom for discriminating different data streams, the system must maintain the number of users satisfying $K\leq 8$.}, as shown in Fig. 5. The results indicate that both the ICGNN and the matrix-inverse-free ICGNN methods (called L-ICGNN) achieve SEs comparable to WMMSE under all configurations, highlighting their universality and effectiveness. Particularly, the L-ICGNN reduces computational complexity by controlling CG method iterations to $T=6 < N_t$, while only causing minimal performance degradation. The WCGCN method outperforms MRT but consistently underperforms compared to ICGNN. This is primarily because, in high-dimensional precoding scenarios, the nonlinear fitting capacity of simple multi-layer fully connected networks is insufficient to fully capture the complex mapping between channel features and precoding vectors. Fig. 6 compares the runtime between WMMSE and ICGNN, showing that ICGNN, with simple forward inference and GPU parallel computing, significantly reduces runtime to about one second. Furthermore, in Table I, we validate ICGNN's generalization ability, with percentages representing performance relative to WMMSE. By training with a base scenario of $K=5$ users, we demonstrate the performance of ICGNN across different numbers of users. The table shows that both the ICGNN and L-ICGNN perform better than WMMSE when extrapolated to larger graphs. There is a slight performance loss on smaller graphs, but they still achieve over 95\% and 98\% performance, respectively.

\begin{figure*}[!t]
\centering
\subfloat[$N_t=2$, $L=7$]{
    \includegraphics[scale=0.55]{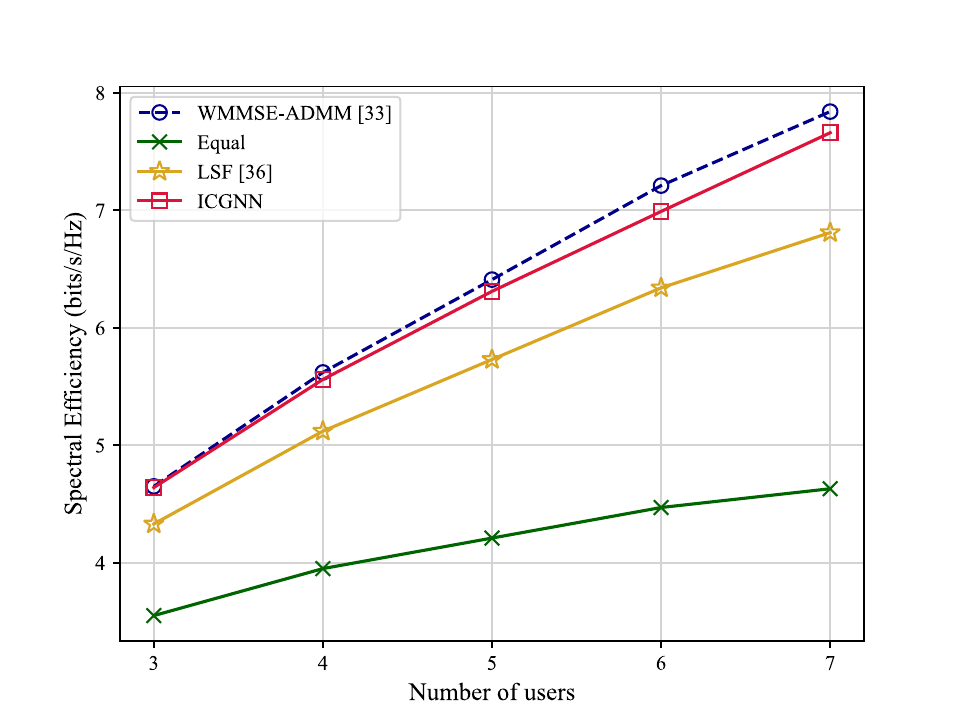}}
\hspace{2em}
\subfloat[$N_t=2$, $L=16$]{
    \includegraphics[scale=0.55]{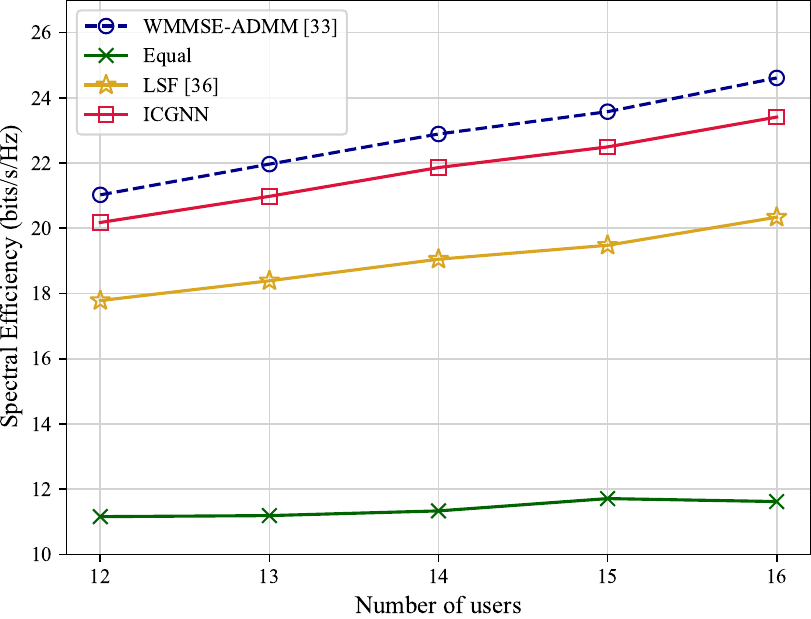}}
\caption{SE versus different numbers of users under two system scales: (a) small-scale system, (b) relatively large-scale system.}
\label{fig:9}
\end{figure*}

\begin{table}[!t]  
\begin{center}   
\caption{SE (bits/s/Hz) of A Model Trained with $N_r=2$ Antennas and Its Generalization to Different Numbers of Users' Antennas.}  
\label{table:2} 
\begin{tabular}{|m{1.34cm}<{\centering}|m{1.1cm}<{\centering}|m{0.6cm}<{\centering}|m{1.8cm}<{\centering}|m{1.8cm}<{\centering}|}   
\hline   \textbf{No. of UEs' Ante.} & \textbf{WMMSE}& \textbf{MRT}& \textbf{ICGNN}& \textbf{L-ICGNN} \\   
\hline   $1$ & $5.18$ & $4.59$ & $4.89(94.4\%)$ & $4.97(95.9\%)$ \\ 
\hline   $2$ & $8.44$ & $7.46$ & $8.47(100.4\%)$ & $8.24(97.6\%)$  \\  
\hline   $3$ & $11.33$ & $9.93$ & $11.24(99.2\%)$ & $10.83(95.6\%)$  \\   
\hline   $4$ & $13.26$ & $11.63$ & $12.93(97.5\%)$ & $12.47(94.0\%)$ \\     
\hline   
\end{tabular}   
\end{center}   
\end{table}

\textit{2)} \textit{Generalization to Different Numbers of User's Antennas:}
We further examine ICGNN's performance with varying numbers of user's antennas in a scenario with $K=3$ users\footnote{Similarly, the number of antennas per user must satisfy $N_r\leq 5$.}. 
As shown in Fig. 7, regardless of the number of antennas per user, both ICGNN and L-ICGNN consistently outperform WCGCN and match WMMSE. This demonstrates the robust versatility of the network model, as it requires no adjustments when dealing with varying numbers of user's antennas. In Table II, we train the model with $N_r=2$ antennas per user and test it under different antenna configurations. The results show that the ICGNN and L-ICGNN continue to exhibit good performance across various configurations, thus confirming its strong generalization ability.

\textit{3)} \textit{Robustness Evaluation under Imperfect CSI:}
In practical scenarios, it is challenging to obtain perfect CSI at the BS due to channel estimation and quantization errors. Therefore, we further evaluate the performance of the proposed algorithm under imperfect CSI conditions. Specifically, we incorporate norm-bounded channel uncertainties into the channel model \cite{csi}, i.e., $\mathbf{H}_k = \hat{\mathbf{H}}_k + \Delta\hat{\mathbf{H}}_k $, where $\hat{\mathbf{H}}_k$ denotes the estimate of $\mathbf{H}_k$ and $\Delta\hat{\mathbf{H}}_k$ represents channel estimation error with satisfying $\|\Delta\hat{\mathbf{H}}_k\|_{\rm F} \leq \varsigma$. As shown in Fig. 8, as the norm-bounded error $\varsigma$ increases, the performance of both the proposed method and the WMMSE algorithm gradually degrades due to the mismatch between the beamformer and the actual channel. Nevertheless, the proposed method consistently achieves better performance than WMMSE, which demonstrates its robustness against channel imperfections. Designing an effective beamforming algorithm to mitigate the negative impact of imperfect CSI remains an important topic for future research.

\begin{figure*}[!t]
\centering
\subfloat[$N_t=2$, $K=3$]{
    \includegraphics[scale=0.55]{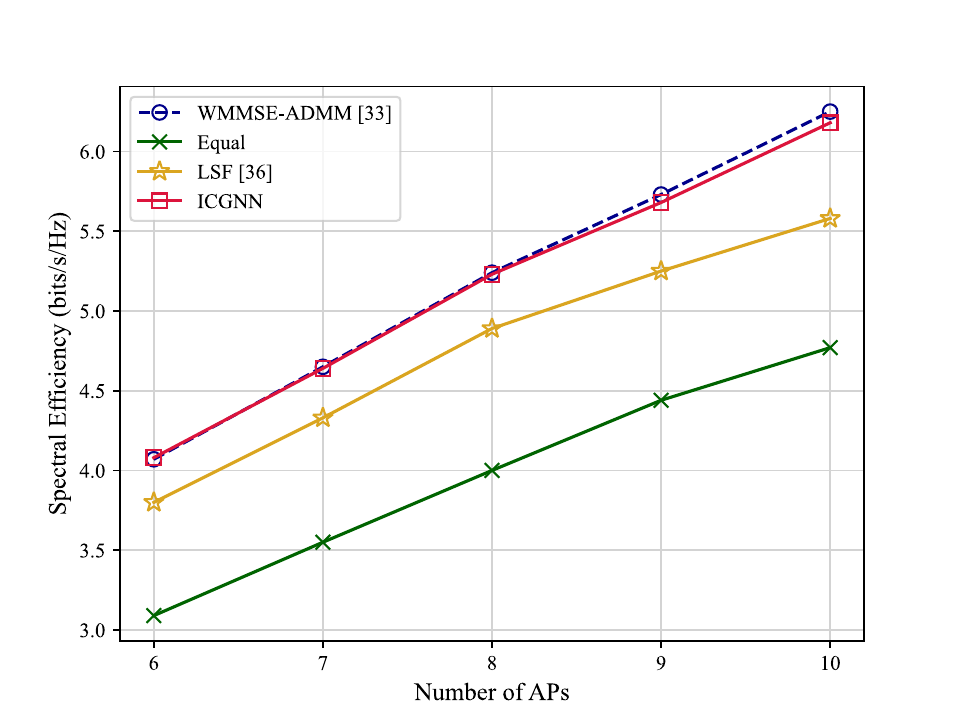}}
\hspace{2em}
\subfloat[$N_t=2$, $K=16$]{
    \includegraphics[scale=0.55]{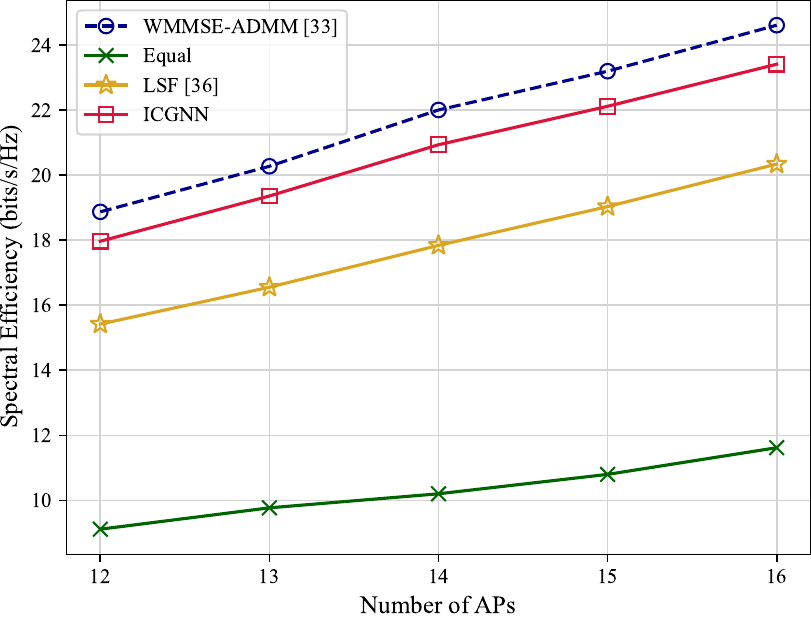}}
\caption{SE versus different numbers of APs under two system scales: (a) small-scale system, (b) relatively large-scale system.}
\label{fig:9}
\vspace{-15pt}
\end{figure*}

\begin{figure*}[!t]
\centering
\subfloat[$L=7$, $K=3$]{
    \includegraphics[scale=0.55]{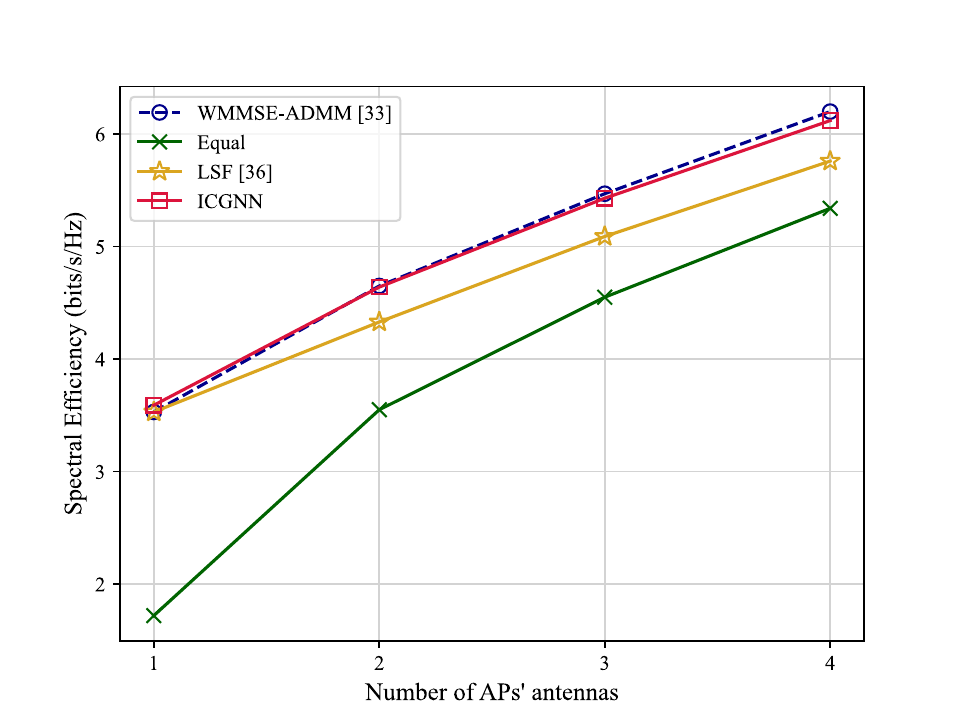}}
\hspace{2em}
\subfloat[$L=16$, $K=16$]{
    \includegraphics[scale=0.55]{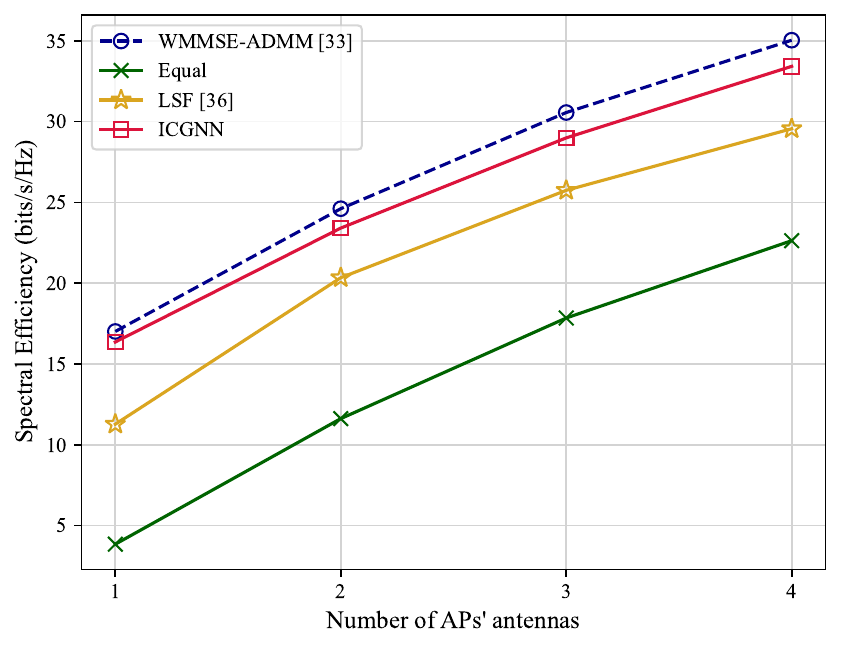}}
\caption{SE versus different numbers of APs' antennas under two system scales: (a) small-scale system, (b) relatively large-scale system.}
\label{fig:10}
\vspace{-15pt}
\end{figure*}

\subsection{Power Allocation for Cell-Free Networks}
In this subsection, we consider a cell-free network where multiple multi-antenna APs and single-antenna users are randomly distributed within a $1000\times1000$ meters area. The transmit power of each AP is $P_t=20$ dBm. The system bandwidth and receiver noise settings are the same as those in subsection A. Each user's uplink transmit power is assumed to be $\rho=10$ mW, and the proportion of downlink data transmission within the entire coherence block is $\tau_d /\tau_c=19/20$. The channel between the APs and users employs a commonly used spatially correlated Rayleigh fading model\cite{ref33}, with large-scale fading still set at $-30.5 - 36.7\log_{10}(d)$ [dB]. Then, we consider the following benchmarks for comparison.
\begin{itemize}
\item{\textit{WMMSE-ADMM}\cite{ref33}: An optimization-based algorithm initialized with equal power allocation and evaluated over 100 iterations with the WMMSE algorithm, using an ADMM penalty factor of $\xi=0.001$.}
\item{\textit{Equal}: This method allocates equal power to all users.}
\item{\textit{LSF}\cite{ref36}: A heuristic approach where the AP allocates power according to the large-scale fading coefficients, using the formula $p_{kl}=P_t\frac{\sqrt{\beta_{kl}}}{\sum_{i=1}^K\sqrt{\beta_{il}}}, \forall k$.}
\end{itemize}


\begin{table}[t] 
\begin{center}   
\caption{SE (bits/s/Hz) of A Model Trained on $K=5$ Users and Its Generalization to Different Numbers of Users.}  
\label{table:3} 
\begin{tabular}{|m{0.9cm}<{\centering}|m{1.2cm}<{\centering}|m{1.1cm}<{\centering}|m{1.1cm}<{\centering}|m{1.8cm}<{\centering}|}   
\hline   \textbf{No. of UEs} & \textbf{WMMSE-ADMM} & \textbf{Equal}& \textbf{LSF}& \textbf{ICGNN}\\   
\hline   $3$ & $4.68$ & $3.57$ & $4.36$ & $4.47(95.5\%)$ \\ 
\hline   $4$ & $5.68$ & $3.98$ & $5.17$ & $5.59(98.4\%)$  \\  
\hline   $5$ & $6.41$ & $4.21$ & $5.73$ & $6.31(98.4\%)$  \\   
\hline   $6$ & $7.16$ & $4.44$ & $6.30$ & $6.98(97.5\%)$ \\     
\hline   $7$ & $7.84$ & $4.63$ & $6.80$ & $7.59(96.8\%)$  \\  
\hline   
\end{tabular}   
\end{center}   
\end{table}

\begin{table}[!t]  
\begin{center}   
\caption{SE (bits/s/Hz) of A Model Trained on $L=8$ APs and Its Generalization to Different Numbers of APs.}  
\label{table:4} 
\begin{tabular}{|m{0.9cm}<{\centering}|m{1.2cm}<{\centering}|m{1.1cm}<{\centering}|m{1.1cm}<{\centering}|m{1.8cm}<{\centering}|}   
\hline   \textbf{No. of APs} & \textbf{WMMSE-ADMM} & \textbf{Equal}& \textbf{LSF}& \textbf{ICGNN}\\   
\hline   $6$ & $4.11$ & $3.12$ & $3.83$ & $4.12(100.2\%)$ \\ 
\hline   $7$ & $4.68$ & $3.57$ & $4.36$ & $4.66(99.6\%)$  \\  
\hline   $8$ & $5.24$ & $4.00$ & $4.89$ & $5.23(99.8\%)$  \\   
\hline   $9$ & $5.66$ & $4.44$ & $5.23$ & $5.61(99.1\%)$ \\     
\hline   $10$ & $6.25$ & $4.77$ & $5.78$ & $6.17(98.7\%)$  \\  
\hline   
\end{tabular}   
\end{center}   
\end{table}
\textit{1)} \textit{Generalization to Different Numbers of Users:}
This subsection investigates a dynamic cell-free network where the number of users serviced by APs varies. Fig. 9 compares power allocation performance between the ICGNN and other benchmark algorithms across varying user counts. Fig. 9(a) and 9(b) illustrate the algorithm's performance under small- and relatively large-scale systems, respectively, to comprehensively validate its effectiveness. 
In both cases, ICGNN consistently achieves performance comparable to the WMMSE-ADMM algorithm, while significantly outperforming heuristic approaches such as LSF and Equal. Under the relatively larger-scale setting, the performance gap between ICGNN and WMMSE-ADMM slightly increases, yet ICGNN still maintains over 95\% of the benchmark's performance, demonstrating its reliability. A potential improvement in ICGNN performance may be realized by adopting a larger neural network architecture, which will be investigated in future work.

To evaluate the generalization capability of the network, we use a small-scale system case as an example. The ICGNN is trained based on a base scenario with $K=5$ users to verify the network's generalization performance across different numbers of users. Simulation results show that this trained model can adapt well to new scenarios with just four gradient descents, achieving over 95\% of WMMSE-ADMM's performance, as demonstrated in Table III. This performance level is unachievable with randomly initialized networks, highlighting ICGNN's strong generalization ability. It has effectively captured the structural information of the underlying graph in the base training scenario, allowing it to adapt to new environments with minimal weight adjustments.

\begin{table}[t]
\begin{center}   
\caption{SE (bits/s/Hz) of A Model Trained on a Base of APs Each Equipped with $N_t=2$ Antennas and Its Generalization to Different Numbers of APs' Antennas.}  
\label{table:5} 
\begin{tabular}{|m{1.4cm}<{\centering}|m{1.2cm}<{\centering}|m{0.9cm}<{\centering}|m{0.8cm}<{\centering}|m{1.8cm}<{\centering}|}   
\hline   \textbf{No. of APs' Ante.} & \textbf{WMMSE-ADMM} & \textbf{Equal}& \textbf{LSF}& \textbf{ICGNN}\\   
\hline   $1$ & $3.59$ & $1.73$ & $3.17$ & $3.60(100.3\%)$ \\ 
\hline   $2$ & $4.65$ & $3.55$ & $4.33$ & $4.64(99.8\%)$  \\  
\hline   $3$ & $5.47$ & $4.54$ & $5.10$ & $5.41(98.9\%)$  \\   
\hline   $4$ & $6.14$ & $5.31$ & $5.72$ & $6.02(98.0\%)$ \\     
\hline   
\end{tabular}   
\end{center}   
\end{table}
\textit{2)} \textit{Generalization to Different Numbers of APs:}
We further explore the dynamic variations in the number of APs serving each user in a cell-free network and evaluate the performance of the ICGNN, as depicted in Fig. 10. The results show that the ICGNN achieves excellent performance across two different system scales, consistently outperforming the LSF and Equal algorithms while matching the performance of the WMMSE-ADMM. Moreover, it exhibits superior performance in small-scale systems.
As demonstrated in Table IV, consider a small-scale system.
We train the model with a base scenario of $L=8$ APs and test its generalization ability with different AP configurations. With just four gradient descents, the ICGNN can fully adapt to new scenarios, consistently achieving over 98.7\% performance, sometimes even surpassing the WMMSE-ADMM algorithm.

\textit{3)} \textit{Generalization to Different Numbers of AP's Antennas:}
As shown in Fig. 11, we evaluate the adaptability of the ICGNN with APs equipped with varying numbers of antennas. It is important to note that our design employs the L-MMSE precoder to reduce the high-dimensional features introduced by increasing AP numbers to a one-dimensional representation, enabling the ICGNN to generalize across different $N_t$ values. However, this approach inevitably results in some feature loss. Therefore, we focus on relatively small variations in the number of AP antennas, which aligns with typical cell-free network configurations where each AP is equipped with a single antenna, and spatial degrees of freedom are provided through multiple AP deployments.
From Fig. 11(a) and 11(b), we observe that the ICGNN achieves performance comparable to the WMMSE-ADMM algorithm, with a slight performance gap observed at the relatively large-scale system. Similarly, to verify its generalization capabilities, we train the model using APs with two antennas in the small-scale case. As shown in Table V, with just four gradient descent iterations, the ICGNN consistently achieves over 98\% performance in new scenarios, sometimes even surpassing the WMMSE-ADMM algorithm, further confirming its robust generalization ability. The negligible time cost of these four gradient descents enables flexible adjustment of training iterations in practical applications.

\begin{figure*}[b]
\vspace{-15pt}
\begin{subequations}
\renewcommand{\theequation}{\theparentequation\alph{equation}}
\hrule 
\vspace{-3pt} 
\begin{align}
(\pi \star\mathbf{\Gamma}^{(l)})_{(\pi(i),:)} &= \mathrm{U}\left(\left((\pi \star\mathbf{X}^{(l-1)})_{(\pi(i),:)}, \underset{j \in \mathcal{N}(\pi(i))}{\operatorname{MAX}}\left\{\mathrm{M}\left(\left((\pi \star\mathbf{X}^{(l-1)})_{(\pi(i),:)}, (\pi \star\mathcal{H})_{(\pi(j), \pi(i),:)}\right)\right)\right\}\right)\right)  \\
(\pi \star\mathbf{\Gamma}^{(l)})_{(\pi(i),:)} &= \mathrm{U}\left(\left(\left((\pi \star\mathbf{X}^{rc(0)})_{(\pi(i),:)}, (\pi \star\mathbf{\Gamma}^{(l-1)})_{(\pi(i),:)}\right), \right.\right. \nonumber \\
&\hspace{4em} \left.\left.\underset{j \in \mathcal{N}(\pi(i))}{\operatorname{MAX}}\left\{\mathrm{M}\left(\left(\left((\pi \star\mathbf{X}^{rc(0)})_{(\pi(j),:)}, (\pi \star\mathbf{\Gamma}^{(l-1)})_{(\pi(j),:)}\right), (\pi \star\mathcal{H})_{(\pi(j), \pi(i),:)}\right)\right)\right\}\right)\right)
\end{align}
\end{subequations}
\end{figure*}
\section{Conclusion}
This paper addressed generalization challenges in deep learning for communications by proposing a universal solution. We modeled wireless networks as virtual homogeneous graphs and used the ICGNN to optimize resource allocation. In cellular networks, the ICGNN effectively handled MU-MIMO precoding with varying user and antenna counts, demonstrating strong generalization. In cell-free networks, we developed a universal power allocation scheme that proved robust against dynamic AP numbers, user counts, and antenna configurations. These results highlighted the ICGNN's potential in complex network environments and its promise for future wireless technologies. In the future, an effective beamformer capable of mitigating the impact of imperfect CSI will be investigated.

\appendices
\section{Proof of the permutation invariance of graph optimization function}
We use mathematical induction to prove the permutation invariance of $g(\cdot,\cdot,\cdot,\cdot)$.

First, for a graph with one node $|V|=1$, the function $g(\cdot,\cdot,\cdot,\cdot)$ depends solely on the node's feature $\gamma_1$ and has no edges. Since a permutation $\pi$ has no effect on a single node, we have $\mathit{g}(\pi \star \alpha_1,\pi \star \boldsymbol{\gamma}_1, \pi \star\mathbf{x}_1^{rc})= \mathit{g}(\alpha_1,\boldsymbol{\gamma}_1, \mathbf{x}_1^{rc})$, for any permutation $\pi$.
Then, assume that for any graph with $n+1$ nodes, if we select any $n$ nodes to form a subgraph, $g(\cdot,\cdot,\cdot,\cdot)$ satisfies:
$\mathit{g}(\pi \star \boldsymbol{\alpha},\pi \star \mathbf{\Gamma}, \pi \star\mathbf{X}^{rc}, \pi \star\mathcal{H})= \mathit{g}(\boldsymbol{\alpha},\mathbf{\Gamma}, \mathbf{X}^{rc}, \mathcal{H})$. 
This holds because $g(\cdot,\cdot,\cdot,\cdot)$ is symmetric, so reordering the inputs does not change the output. For example, with the sum function $f(\boldsymbol{\alpha})=\sum_{i=1}^{n}\alpha_i$, $f(\pi \star \boldsymbol{\alpha})=\sum_{\pi(i)=1}^{n}\alpha_{\pi(i)}=\sum_{i=1}^{n}\alpha_i$ still remains true. 

Next, consider the graph with $n+1$ nodes. For a new node $v_{n+1}$, and the corresponding weight factor $\boldsymbol{\alpha}'$, target feature matrix $\mathbf{\Gamma}'$, channel feature matrix $\mathbf{X}^{rc'}$, and edge feature tensor $\mathcal{H}'$, we need to prove that
$\mathit{g}(\pi \star \boldsymbol{\alpha}',\pi \star \mathbf{\Gamma}', \pi \star\mathbf{X}^{rc'}, \pi \star\mathcal{H}')= \mathit{g}(\boldsymbol{\alpha}',\mathbf{\Gamma}', \mathbf{X}^{rc'}, \mathcal{H}')$.
According to the induction hypothesis, the permutation invariance holds for any selected $n$ nodes; therefore, the new objective function $\mathit{g}(\boldsymbol{\alpha}',\mathbf{\Gamma}', \mathbf{X}^{rc'}, \mathcal{H}')$, now including $n+1$ nodes, retains its permutation invariance. Therefore, by induction, the permutation invariance of the objective function $g(\cdot,\cdot,\cdot,\cdot)$ holds for any number of nodes $n$.

Similarly, following the proof above, it can be easily shown that the constraint function $C(\cdot,\cdot,\cdot)$ also possesses permutation invariance.

\section{Proof of the permutation invariance in ICGNN}
Taking the feature update process of node $v_i$ as an example, based on the graph isomorphism properties, we have
$(\pi \star \boldsymbol{\alpha})_{\pi(i)}= \alpha_{i}$,
$(\pi \star\mathbf{X}^{(0)})_{(\pi(i),:)}=\mathbf{X}_{(i, :)}^{(0)}$,  
$(\pi \star\mathcal{H})_{(\pi(j), \pi(i),:)}=\mathcal{H}_{(j, i, :)}$, and  
$\mathcal{N}(\pi(i))=\{\pi(j), j \in \mathcal{N}(i)\}$.

Given that the $\operatorname{MAX}(\cdot)$ function is permutation invariant, meaning $\operatorname{MAX}(\pi \star \mathbf{X})=\operatorname{MAX}(\mathbf{X})$, the output of the $l$-th layer for any permuted node $i$, as indicated in equations (30a) and (30b), leads us to the following conclusion. We can thus determine that the updated node features at the $l$-th layer are as described below:
\begin{equation}
\label{eqa:30}
\begin{aligned}
(\pi \star\mathbf{X}^{(l)})_{(\pi(i),:)} &= \left((\pi \star\mathbf{X}^{rc(0)})_{(\pi(i),:)}, (\pi \star\mathbf{\Gamma}^{(l)})_{(\pi(i),:)}\right).
\end{aligned}
\end{equation}

Since this process is valid for any node $i$ and any layer $l$, we deduce that $\psi\left(\pi \star \mathbf{X}^{(l-1)}, \pi \star \mathcal{H}\right)=\pi \star \mathbf{X}^{(l)}$. By applying this recursive relationship through a pipeline workflow across $L$ layers, we infer that $\psi\left(\pi \star \mathbf{X}^{(0)}, \pi \star \mathcal{H}\right)=\pi \star \mathbf{X}^{(L)}$. This outcome demonstrates that the function $\psi(\cdot)$ exhibits permutation equivariance.

Given that a permutation-equivariant function nested within a permutation-invariant function results in a permutation-invariant function, this characteristic leads to $g\left(\pi \star \boldsymbol{\alpha},\psi\left(\pi \star \mathbf{X}^{(0)}, \pi \star \mathcal{H}\right),\pi \star \mathcal{H}\right)=g\left(\pi \star \boldsymbol{\alpha},\pi \star \mathbf{X}^{(L)}, \pi \star \mathcal{H}\right)=g\left(\boldsymbol{\alpha},\mathbf{X}^{(L)}, \mathcal{H}\right)=g\left(\boldsymbol{\alpha},\psi\left(\mathbf{X}^{(0)}, \mathcal{H}\right), \mathcal{H}\right)$. This completes the proof for Proposition 3.

\end{document}